\documentclass[varenna]{cimento}
\newcommand{\Rs}{R_{sub}}
\newcommand{\Rt}{R_{tot}}

\newcommand{\alf}{Alfv\'en }
\newcommand{\dHybridR}{{\tt dHybridR}}

\newcommand{\vsh}{v_{\rm sh}}
\newcommand{\rt}{R_{\rm tot}}
\newcommand{\trt}{\Tilde{R}_{\rm tot}}

\newcommand{\rs}{R_{\rm sub}}

\newcommand{\w}[1]{v_{A,#1}}

\newcommand{\de}{\partial}
\newcommand{\omci}{\omega_c^{-1}}

\newcommand\thbn{\vartheta}
\renewcommand{\deg}{^{\circ}}
\newcommand{\xcr}{\xi_{\rm c}}
\newcommand{\esh}{E_{\rm sh}}
\newcommand{\Em}{E_{\rm max}}

\usepackage{graphicx}
\usepackage{amssymb}
\usepackage{amsmath}
\usepackage{wrapfig}

\title{Particle Acceleration at Shocks: An Introduction}

\author{D.Caprioli}

\institute{Department of Astronomy \& Astrophysics and
E. Fermi Institute\\ The University of Chicago, 5640 S Ellis Ave, Chicago, IL 60637, USA}

\begin{document}

\maketitle

\begin{abstract}
These notes present the fundamentals of Fermi acceleration at shocks, with a special attention to the role that supernova remnants have in producing Galactic cosmic rays.
Then, the recent discoveries in the theory of diffusive shock acceleration (DSA) that stem from first-principle kinetic plasma simulations are discussed.
When ion acceleration is efficient, the back-reaction of non-thermal particles and self-generated magnetic fields becomes prominent and leads to both enhanced shock compression and particle spectra significantly softer than those predicted by the standard test-particle DSA theory. 
These results are discussed in the context of the non-thermal phenomenology of astrophysical shocks, with a special focus on the remnant of SN1006.
\end{abstract}

%\section{Acceleration of galactic cosmic rays}
\section{The SNR Paradigm for the Origin of Galactic CRs}
The origin of cosmic rays (CRs) has been an outstanding issue in Astrophysics since the pioneering discovery by V. Hess in 1911. 
At least for relatively low-energies, below and around the so-called knee of the overall CR spectrum ($\sim 10^{15}$ eV), the best source candidates have been supernova remnants (SNRs).

In 1934 Baade and Zwicky  \cite{baade+34} suggested that supernova (SN) explosions were due to the release of a huge amount of gravitational binding energy during the transition from an ordinary star to a \emph{neutron star};
not satisfied of having introduced this groundbreaking idea, they also argued that SNe were also responsible for the acceleration of CRs.
The latter statement was based on an energetic argument which estimated a fraction of 20--30 per cent of the SN ejecta kinetic energy ($\sim 10^{51}$ erg) to be channelled into relativistic particles in order to account for the CR energetics.
Despite Baade and Zwicky were concerned with extragalactic SNe, the idea that Galactic CRs are accelerated in Galactic SNRs is widely popular and usually referred to as the SNR paradigm 
\cite{hillas05}.

Already in the '60s, Ginzburg and Syrovatskii interpreted the radio emission of many SNRs as a result of synchrotron radiation emitted by accelerated electrons \cite{ginzburg-syrovatskii68}, providing the first observational evidence that particle acceleration should be effective in SNRs. 
Only with the advent of gamma-ray astronomy, though, it has been possible to infer that such systems can efficiently accelerate protons and heavier nuclei, as attested by the GeV and TeV photons that come from the decay of neutral pions produced by nuclear interactions between CRs and thermal plasma \cite{morlino+12,ackermann+13}.

An energetic argument alone, however, cannot be a satisfactory theory of particle acceleration; 
in the past century, much effort has gone into figuring out the exact process through which a few particles can be extracted from a thermal plasma and be accelerated to ultra-relativistic energies, with spectra that are extremely regular, usually fitted with power-law distribution in momentum.
In the past few decades, the advancement of numerical methods and the advent of modern supercomputers have made possible \emph{ab-initio} simulations of astrophysical plasmas and allowed astrophysicists to model the complex, non-linear, interplay between particles and electromagnetic waves, eventually unravelling the processes responsible for the energization of the highest-energy particles in the Universe.

In this lecture we present the basic ingredients of the theory of particle acceleration in collisionless shocks, focusing on the basic properties of Fermi acceleration and diffusive shock acceleration (DSA). 
In particular, we outline the recent progress in modeling DSA via kinetic plasma simulations, especially hybrid simulations (with kinetic ions and fluid electrons), and discuss the observational counterparts of such findings.

\section{Acceleration at Shocks}\label{sec:intro}
In this section we lay the groundwork for the theory of self-sustained particle acceleration at shocks. 
More precisely, we discuss how repeated interactions of a particle with the magnetic structures embedded in a shock may lead to energy gain and to universal spectra of accelerated particles.

\subsection{Magnetic Mirroring}
When a magnetic field is effectively constant over one Larmor gyration, a particle conserves its magnetic moment $\mu=p_\perp^2/B$, where $p_\perp$ is the component of the momentum transverse to the magnetic field. 
Since magnetic fields do no work, the total momentum $p^2= p_\parallel^2 + p^2_\perp$ is constant, too. 
This means that if a particle enters a region with stronger $B$, its $p_\perp$ must increase, and hence its $p_\parallel$ decreases;
if the magnetic gradient is sufficiently large, the particle may come to a stop before eventually reversing its motion.
This effect is known as \emph{magnetic mirroring} and can be regarded as an effective way of scattering particles in non-homogeneous magnetic fields.

%\subsection{The Fermi mechanism --- Full derivation}

\subsection{The Fermi Mechanism}\label{ssub:fermimec}
Soon after listening to a lecture at the University of Chicago, in which  by H.~\alf  argued that the interstellar medium is permeated by magnetic irregularities, E.~Fermi realized that such scattering of particles against such \alf waves could naturally lead to energization \cite{fermi49,fermi54}. 
In fact, collisions are elastic in the mirror frame, but 
when mirrors move, they may lead either to an energy gain if they are head-on, or to an energy loss, if they are tail-on;
physically speaking, an acceleration arises because of the motional electric field.

In astrophysical contests, where the mirrors are typically represented by Alfv\'en waves or by magnetized clouds (henceforth, magnetic irregularities), particles may \emph{statistically} be accelerated as a consequence of the fact that head-on collisions are more frequent than tail-on ones\footnote{One may ponder the analogy with inverse-Compton scattering: Fermi, who moved to Chicago invited by A.~Compton himself, must have been quite familiar with the idea...}.
If a thermal particle succeeds in crossing an \emph{injection}  threshold \cite{caprioli+15}, which may be dictated either by ionization losses or by the minimum energy to enter the acceleration process, it can achieve relativistic energies and become a bona-fide CR.

Let us consider the interaction between a mirror that moves with velocity $\vec{V}$ along the $\hat{x}$-axis and a particle moving with velocity $\vec{v}=(-v \cos\vartheta,-v\sin\vartheta)$, such that $\vartheta=0$ corresponds to a head-on collision.

Posing $c=1$ for simplicity, the initial energy and momentum of such a particle be $E_i$ and $p_i=v E_i$.
Performing a Lorentz transformation into the mirror frame, where quantities are labelled with the prime subscript, we have
\begin{equation}
    E'_i=\Gamma (V p_{xi} +E_i); \quad p'_{xi}=\Gamma (p_{xi}+VE_i),
\end{equation}
where $\Gamma=1/\sqrt{1-V^2}$ is the Lorentz factor of the boost.
If the reflection is elastic in the mirror frame, then the final energy and momentum along $x$ are
\begin{equation}
    E'_f=E'_i;\quad p'_{xf}=-p'_{xi}.
\end{equation}
Boosting back in the laboratory frame one has
\begin{equation}\label{energygain}
    E_f= \Gamma^2 E_i (2 V v \cos\vartheta +1 + V^2), 
\end{equation}
so that the fractional energy gain in a reflection is
\begin{equation}\label{eq:de}
    \frac{\Delta E}{E}\equiv\frac{E_f-E_i}{E_i}=2 V (v\cos\vartheta+V).
\end{equation}

In general, the rate of collisions is proportional to the relative velocity between the mirror and the particle.
The projection of the mirror's speed along $\vec{v}$ is $V\cos\vartheta$; 
using the relativistic composition of velocities, the probability of collision per solid angle reads (consider the symmetry in the azimuthal angle) is:
\begin{equation}\label{angdistrib}
    dP(\cos\vartheta)\propto \frac{V\cos\vartheta+v}{1+vV\cos\vartheta} {\rm d}\cos\vartheta\propto (1+V\cos\vartheta)d\cos\vartheta,
\end{equation}
where in the last step we also assumed $v\to 1$ and $V\ll 1$ (relativistic particle scattered by a non-relativistic mirror).
Note that Eq.~\ref{angdistrib} contains an important piece of physics: it expresses the fact that head-on collisions ($\cos\vartheta=1$)  are more probable than tail-on ones ($\cos\vartheta=-1$), which is at the basis of Fermi acceleration.
Finally, averaging the energy gain in Eq.~\ref{eq:de} over an isotropic distribution and reintroducing $c$, one gets:
\begin{equation}\label{IIOFM}
\left\langle  \frac{\Delta E}{E} \right\rangle_P=\frac{8}{3}\frac{V^2}{c^2}.
\end{equation}
Since the gain is  $\propto (V/c)^2$, this process is often referred to as \emph{second-order Fermi mechanism}, and  may be not very efficient since typically $V\ll c$; for instance, for interstellar magnetic fluctuations $V$ is of the order of the \alf speed, i.e., $\sim 1-10$ km/s, so that $\sim 10^{10}$ collisions would be needed just to double the particle's energy.
This is where Fermi's original idea of accelerating CRs in the interstellar turbulence  \emph{quantitatively} fails: it is just too slow (in addition to not providing universal spectra, as we will show below).

If all the collisions were head-on, though, the average over the distribution in Eq.~\ref{angdistrib} would be only on values of $\cos\vartheta>0$, and the energy gain would be dominated by the first term in Eq.~\ref{energygain}, which is instead proportional to $V/c$ (\emph{first-order Fermi mechanism}). 

In the late '70s, various authors independently realized that, when Fermi mechanism is applied to shocks, it can lead to a very efficient acceleration of CRs \cite{krymskii77,axford+78, blandford+78, bell78a, bell78b}.
This process, referred to as \emph{Diffusive Shock Acceleration} (DSA), is based on the fact that ---if particles can be repeatedly scattered back and forth across the shock--- they can experience multiple head-on collisions and gain energy very efficiently.

Before diving into the details of how particles can be accelerated, it is worth sketching a brief summary of the main characteristics of shocks.

\subsection{Shock Hydrodynamics}\label{ssec:hydro}
\begin{wrapfigure}[11]{r}{0.47\textwidth}
\centering
\vspace{-1cm}
\hspace{0.3cm}
		\includegraphics[width=0.45\textwidth]{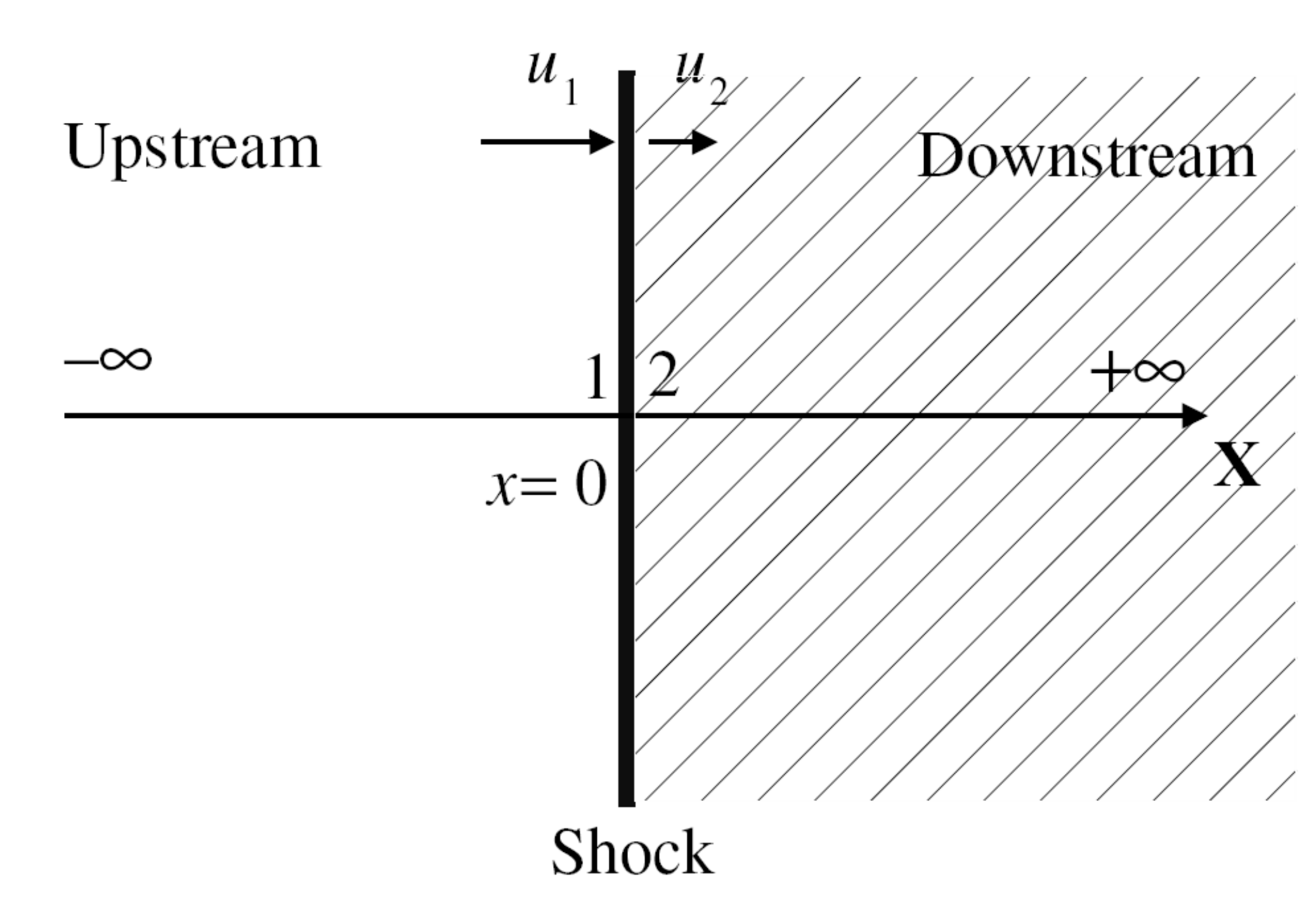}
%  \vspace{-.3cm}
\caption{Schematic structure of a shock, in a frame centered on the discontinuity: the upstream medium is un-shocked and moves towards the shock with velocity $u_1=\vsh$.}
	\label{fig:shockscheme}
\end{wrapfigure}

The hydrodynamics of a stationary, non-viscous, non-relativistic fluid is described by the classical equations for the conservation of mass, momentum and energy \cite{landau6}. 
A 1D shock is a solution of such equations where there is a surface across which physical quantities are discontinuous and entropy increases. 
In the shock frame, where the system can be assumed to be stationary, in the sense that quantities on both sides of the shock are homogeneous and time-independent, the differential conservation laws for mass, momentum and energy read:
\begin{equation}
\frac{\partial}{\partial x}\left( \rho u \right) = 0\,;
\end{equation}
\begin{equation}
\frac{\partial}{\partial x} \left( \rho u^2 + p\right)=0\,;
\end{equation}
\begin{equation}
\frac{\partial}{\partial x} \left( \frac{1}{2}\rho u^3 + 
\frac{\gamma }{\gamma-1}p u\right) = 0\,.
\end{equation}
As usual, $\rho$, $u$, $p$ and $\gamma$ stand for density, velocity, pressure and ratio of the specific heats of the fluid (i.e., the adiabatic index). 
In the two terms of the energy conservation equation, we recognize the kinetic energy flux and the enthalpy flux.

In this simple treatment each physical quantity is continuous both ahead (\emph{upstream}) and behind (\emph{downstream}) the shock, but may be discontinuous at the shock surface. 
The conditions which describe the jumps of these quantities between the shocked (labelled with the subscript 2) and unshocked physical quantities (labelled with subscript 1) are known as \emph{Rankine--Hugoniot relations}:
\begin{equation}
R\equiv\frac{\rho_2}{\rho_1}=\frac{u_1}{u_2}=\frac{(\gamma+1)M_1^2}{(\gamma-1)M_1^2+2}
\to \frac{\gamma+1}{\gamma-1}\;,
\end{equation}
\begin{equation}
	\frac{p_2}{p_1}=\frac{2\gamma M_1^2-\gamma+1}{\gamma+1}
 \to \frac{2\gamma M_1^2}{\gamma+1}\;,
\end{equation}
\begin{equation}
	\frac{T_2}{T_1}=\frac{[2\gamma M_1^2-(\gamma-1)][(\gamma-1)M_1^2+2]}{(\gamma+1)^2M_1^2}
 \to 2\gamma\frac{\gamma-1}{(\gamma+1)^2}M_1^2\;,
\end{equation}
where we introduced the compression factor between the downstream and upstream density $R\equiv\rho_2/\rho_1$ and the sonic Mach number $M_1\equiv u/c_{s,1}$, where the sound speed $c_s$ is defined as $c_s^2= \gamma p/\rho$.
The asymptotic values for $M_1^2\gg 1$ are also reported after the $\to$ symbol. 
For an ideal monoatomic gas $\gamma=5/3$, so that generally $R=4$ for any strong shock. 
We notice that both the pressure and temperature jumps are instead proportional to $M_1^2$, so that a strong shock tends to heat up the downstream plasma very efficiently. 
Astrophysical shocks are often very strong, so we expect the shock dynamics to often be in such a regime.

For the Rankine--Hugoniot conditions of shocks propagating into a large-scale magnetic field, in the MHD limit, one may refer to \cite{jones+91, treumann09}, while the relativistic generalization is provided by the Taub conditions \cite{taub48}.

\subsection{Diffusive Shock Acceleration}\label{tpDSA}
As mentioned above, DSA was proposed at the same time out by several authors \cite{krymskii77,axford+78, blandford+78, bell78a, bell78b} as a way to achieve a first-order Fermi acceleration, with the additional bonus that the velocity of the ``mirror'' could be very large ($V\approx$ thousands of km/s for SNR shocks).

Let us consider a shock, in its reference frame where the downstream (upstream) is for $x>0(x<0)$, so that both $u_1$ and $u_2$ are positive (see figure \ref{fig:shockscheme}).
Let us also follow a test particle that is initially in the downstream with velocity $v\gg u_1$ and flight cosine $\mu=\cos\vartheta$; 
in this situation, $V=u_1-u_2$ is the relative velocity between upstream and downstream fluids.
If the particle crosses the shock, its energy in the upstream frame would read
\begin{equation}
    E'_i=\Gamma E_i (1+V v \mu_i),
\end{equation}
where we used the same notation as above.
While upstream, the particle may be scattered back by magnetic turbulence and re-encounter the shock with a different direction $\mu'$ and velocity $v'$.
If the particle makes it back downstream, a cycle is completed and its final energy reads: 
\begin{equation}
    E_f=\Gamma^2 E_i (1+Vv \mu_i)(1-Vv'\mu').
\end{equation}
We need to express all the directions in the downstream frame, so we use the relativistic transformation of angles
\begin{equation}
    \mu_f=\frac{\mu'v-V}{1-\mu'V} \to \mu'=\frac{\mu_f+V}{v+\mu_fV}.
\end{equation}
Finally, assuming $v\to 1$ and $v'\to 1$ (relativistic particles), we obtain 
\begin{equation}\label{eq:efei}
    \frac{E_f}{E_i}\simeq \frac{1+V\mu_i}{1+V\mu_f}.
\end{equation}
This equation holds for \emph{arbitrary} shock speed $V\leq 1$ and conveys the fact that the energy gain only depends on the in/out-going directions. 
For relativistic shocks, this leads to a typical energy gain of the order of $\Gamma^2$, analog to a Compton scattering \cite{vietri03}.
Such an energy gain may be important also for one-shot interactions of CRs with relativistic jets (\emph{espresso} mechanism), which is potentially important for the acceleration of ultra-high-energy CRs and neutrinos in relativistic AGN jets \cite{caprioli15,mbarek+22}.

Here we focus on the non-relativistic limit $V\ll 1$, which returns
\begin{equation}
    \frac{E_f}{E_i}\simeq (1+V\mu_i)(1-V\mu_f).
\end{equation}
As we did before for the mirror-particle scattering, we want to calculate the mean fractional energy gain, averaging on both the initial and final distributions of $\mu$.

In order to do this, we need to calculate the directions $\mu$ that lead to a shock crossing.
More precisely, for the particle to cross from downstream to upstream one needs 
\begin{equation}
    \mu_i+u_2\leq 0 \to -1\leq \mu_i\leq -u_2,
\end{equation}
while for crossing from upstream to downstream one requires
\begin{equation}
    \mu_f+u_2\geq 0 \to -u_2\leq \mu_f \leq 1.
\end{equation}
Assuming that particles are isotropized in both the upstream and downstream reference frame, the probability of interacting is proportional to the particle flux in the direction $x$, i.e., $P(\mu)\propto\mu$ (see Eq.~\ref{angdistrib}).
Performing the average of Eq.~\ref{eq:efei} with such a probability and for the extremes of integration above, one obtains (after some simple but somewhat tedious algebra):
\begin{equation}\label{studio}
    \left\langle \frac{\Delta E}{E}\right\rangle\simeq \frac{4}{3}\frac{u_1-u_2}{c},
\end{equation}
which expresses the fractional energy gain for one DSA cycle at a non-relativistic shock.

The presence of a tangled magnetic field, or of any kind of spatial diffusion, is fundamental to ensure that the process may occur several times.
Strong particle diffusion may not be sufficient if particles are not \emph{injected} into the acceleration process, as it is clear that not all of the thermal particles can be promoted to CRs.
The exact conditions required for ions and electrons to be injected into DSA will be discussed below, but at this stage it is worth pointing out that \emph{DSA is not bound to happen at any shock}.
For instance, in a relativistic shock one has $u_2=c/3$ and it may be very hard for particles to swim back upstream once they have crossed the shock.
If the magnetic geometry of the shock allowed it (see, e.g., the discussion in \cite{sironi+09}), in its first cycle a particle could gain energy proportional to $\Gamma^2$; 
yet, multiple iterations of the process would be severely discouraged and/or yield much smaller energy gain due to the correlation between $\mu_i$ and $\mu_f$ \cite{achterberg+01}.
For a detailed treatment of these aspects of particle acceleration at relativistic shocks see, e.g., refs.~\cite{vietri03,blasi+05,morlino+07a,morlino+07b}.

\subsection{Bell's Argument for Power-law Distributions}
The most important property of DSA is that it accelerates particles with a spectrum that is fully independent of the details of particle scattering. 
This can be shown with Bell's elegant approach \cite{bell78a}. 
Let us consider $N_0$ test particles of initial energy $E_0$ injected in a generic acceleration mechanism; 
be $G=\Delta E/E$ the fractional energy gain per cycle and  $1-P$ the probability of leaving the accelerator after each cycle. 
%Particles can not be endless accelerated because they are clutched to the frozen-in magnetic field lines, hence at some point they have to be advected with the downstream fluid.
%In particular, while upstream advection forces particles to return at the shock, downstream it pulls them away from the shock with roughly constant velocity $u_2$, corresponding to a length scale of order $u_2 t$. 
%The diffusion due to magnetic irregularities, instead, makes particle travel for typical random-walk scale lengths, which are proportional to $t^{1/2}$: it is evident that the return from downstream is allowed only for small $t$, more precisely until the diffusion length remains comparable with the advection length scale.
%If $E_0$ is the initial energy of the set of $N_0$ particles and G is the energy gain given by \ref{studio} 
%\begin{equation}
%	G\equiv\frac{E_1}{E_0}=1+\frac{4}{3}\frac{u_1-u_2}{c}\,,
%\end{equation}
After one cycle there will be $N_0 P$ particles with energy $GE_0$
Similarly, after $k$ steps there will be $N_k=N_0P^k$ particles with energy $E_k=E_0G^k$. 
Therefore, eliminating $k= \ln{(N_k/N_0)/\ln P}=\ln{(E_k/E_0)/\ln G}$, we obtain
\begin{equation}\label{adef}	\ln\left(\frac{N_k}{N_0}\right)=Q\ln\left(\frac{E_k}{E_0}\right)\to 
N_k = N_0\left(\frac{E_k}{E_0}\right)^{-Q}
\qquad Q\equiv-\frac{\ln P}{\ln G}.
\end{equation}

For a shock, $G$ is given by Eq.~\ref{studio}, while the probability of leaving the accelerator corresponds to the probability of being advected away downstream, i.e, $P=J_-/J_+$ is the ratio between the flux of particles returning to the shock, $J_-$, and the flux of particles impinging on the shock, $J_+$ (see Fig.~\ref{fig:shockscheme}).
Calling $J_\infty$ the flux of particles escaping towards downstream infinity, stationarity requires
\begin{equation}
	J_+=J_-+J_\infty~.
\end{equation}
If $n$ is the particle number density and particles are isotropic in the fluid frame, then
\begin{equation}	J_+=\int^{1}_{0}d\cos\vartheta\frac{nc}{2}\cos\vartheta=\frac{nc}{4}~,
\end{equation}
while the flux advected to infinity is, by definition,
\begin{equation}
	J_\infty=nu_2~
\end{equation}
and eventually:
\begin{equation}\label{pi}	P=\frac{J_+-J_\infty}{J_+}=\frac{c}{c+4u_2}\approx1-\frac{4u_2}{c}~,
\end{equation}
where the last approximation holds for non-relativistic shocks, i.e., $u_1,u_2\ll c$.

In the same limit, one can substitute Eq.~\ref{pi} in Eq.~\ref{adef} and obtain
\begin{equation}\label{eq:bell}
	Q=-\frac{\ln P}{\ln G}\approx \frac{-\ln(1-4u_2/c)}{\ln{[1+4/3(u_1-u_2)/c]}}\approx \frac{3u_2}{u_1-u_2}\approx\frac{3}{R-1}\,.
\end{equation}

Note that $Q$ corresponds to the spectral index of the \emph{integral} energy spectrum (i.e., the total number of particles with energy larger than $E$);  
the \emph{differential} spectrum then reads:
\begin{equation}\label{specslope}
	\frac{dN(E)}{dE}=f(E)\propto E^{-q_E}\quad {\rm with}\quad
	q_E=Q+1=\frac{R+2}{R-1}\,.
\end{equation}
It is very important to notice that the spectral index $q_E$ depends only on the compression ratio $R=u_1/u_2$.
This means that for \emph{any} strong shock, i.e.\ a shock for which $V/c_s\gg 1$ and consequently $R=4$, one always gets $q_E=2$, independently of the details of the mechanisms responsible for the diffusion.

Note that the derivation above holds for relativistic particles at non-relativistic shocks.
If one relaxes the hypothesis that particles are relativistic, the universal spectrum is a power-law in momentum that reads:
\begin{equation}\label{eq:DSA}
    \frac{dN(p)}{dp}=4\pi p^2 f(p)\propto p^{-q_p}\quad {\rm with}\quad q_p=\frac{3R}{R-1}\,.
\end{equation}
The conversion between energy and momentum must be done remembering that 
\begin{equation}
4\pi p^2 f(p)dp= f(E)dE \to f(E)=4\pi p^2 f(p)\frac{dp}{dE};
\end{equation}
therefore, for $R=4$, in the non-relativistic regime $E=p^2/2m$ and $q_E=1.5$, while 
in the relativistic limit $E\propto p$ and $q_E=2$.

At relativistic shocks \cite{blasi-vietri05,morlino+07a}, or when shocks are strongly magnetized and the magnetic field is quasi-perpendicular to the shock normal \cite{kirk+96,bell+13}, instead, the assumptions that the CR distribution is isotropic in the fluid frame should be relaxed, in general leading to spectra that deviate from the universal one and depend on the details of the scattering process.
Also, when CR acceleration is efficient, we anticipate that magnetic fluctuations may have a finite speed $u_w$ with respect to the thermal gas, so that the effective compression ratio seen by the particles may differ from $R$; this effect is discussed in \S\ref{sec:modified}.

\subsection{A Kinetic Derivation}
It is instructive to derive the results above also from a kinetic perspective, i.e., by solving the Vlasov (also known as the non-collisional Boltzmann) equation for the CR distribution function $f(\vec{x},\vec{p},t)$:
\begin{equation}
	\frac{\de f}{\de t}+\vec v(p)\cdot \vec{\nabla}_x f+\frac{\de \vec p}{\de t}\cdot \vec{\nabla}_p f=0\qquad\frac{\de \vec p}{\de t}=q\left[\vec E+\frac{1}{c}\vec v(p)\times\vec B\right]\;.
\end{equation}
The standard approach is to introduce some assumptions that allow reducing this equation to a transport equation that can be solved even analytically when the shock structure is known \cite{skilling75a,skilling75b,skilling75c,amato+05,caprioli+09a,caprioli+10a,caprioli+10c,caprioli12,diesing+21}.
Such main assumptions are:
1) CRs have Larmor radii larger than the shock thickness (i.e., the equation does not apply to thermal particles) and in general $v\gg u$;
2) CRs undergo many small pitch-angle scatterings on magnetic irregularities (generically called \emph{waves}), which drive them towards isotropy in the local wave frame;
3) waves propagate parallel to the background field $\vec{B}_0$, assumed to be \emph{aligned} with the shock velocity, i.e., we have a \emph{parallel shock}\footnote{When $\vec{B}_0$ is perpendicular to the shock normal, we talk of a \emph{perpendicular shock.}}.
Under these assumptions, the CR distribution function can be expanded, in the wave frame, as a sum $f=f_0+f_1+...$ of terms with growing anisotropy, with the $n^{\rm th}$ term $\propto\mathcal{O}(u/c)$. 
For non-relativistic shocks, such an expansion can be truncated at the second order and the first-order term can be written as a diffusive flux, proportional to the spatial derivative of $f_0$.
Detailed calculation of the procedure described above can be found for example in ref.~\cite{skilling75a} or in the textbook at ref.~\cite{vietri}. 

In the case of a non-relativistic, one-dimensional, parallel shock, we obtain a \emph{diffusion--advection equation} for the CR isotropic part of the distribution function $f_0=f(x,p)$:
\begin{equation}\label{diff-conv}
	\frac{\de f(x,p)}{\de t}+\Tilde{u}\frac{\de f(x,p)}{\de x}=\frac{\de}{\de x}\left[D(x,p)\frac{\de f(x,p)}{\de x}\right]+
	\frac{p}{3}\frac{d\Tilde{u}}{dx}\frac{\de f(x,p)}{\de p}\,,
\end{equation}
Here we have introduced the effective fluid speed $\Tilde{u}=u+v_w$, which accounts for a finite velocity of the scattering centers (waves) with respect to the fluid, which can be both positive and negative, and the \emph{diffusion coefficient} $D(x,p)$, which includes all the details of the interaction between the waves and the particles and describes the spatial random-walk of particles along the gradient of $f$.
The four terms correspond to: time evolution, advection, diffusion, and adiabatic compression, respectively.

Using the same notation as Fig.~\ref{fig:shockscheme} and assuming that: 
1) $f$ is continuous at the shock, which comes from the fact that CRs have large Larmor radii; 
2) $f(x\to\infty)\to 0$, i.e., no CRs far upstream;
3) $f$ is homogeneous downstream, i.e., $\de f/\de x(x>0)=0 $, which is the only acceptable choice in the stationary limit; 
and 
4) $D$ is independent of $x$ upstream (downstream it does not matter because of assumption 3),
we can promptly solve Eq.~\ref{diff-conv} obtaining:
\begin{equation}
    f(x,p)= f_s(p)\exp\left(\frac{ux}{D} \right); \quad f_s(p)\propto p^{-q_p},\quad  q_p=\frac{3\Tilde{R}}{\Tilde{R}-1},
\end{equation}
where $f_s(p)$ is the CR distribution at the shock and $q_p$ is the same derived in Eq.~\ref{eq:DSA}, provided that the velocity of the scattering centers is small with respect to the fluid speed.
Since typically $v_w$ is of the order of the \alf speed, this is usually thought to be a good approximation, but we will see below that this is not generic and has important implications.
Again, we find that the CR spectrum is a power law in momentum, which depends only on the shock compression ratio (better, on the compression ratio of the scattering center velocity $\Tilde{R}=\Tilde{u_1}/\Tilde{u_2}$).
%\begin{equation}
%    \phi(x,p)=\Tilde{u}f -D \frac{\de f}{\de x} = {\rm constant}=\phi_\infty =0,
%\end{equation}
%i.e., the advective flux balances the diffusive flux (the last step comes from assuming that $f(x\to\infty)\to 0$, i.e., no CRs at upstream infinity).

In summary, the very reason why DSA is so attractive is that it just depends on particles tending to become isotropic in the fluid (or wave) frame on each side of the shock; 
since there is a relative velocity $u_1-u_2$ between the two frames, repeatedly achieving such an isotropization requires energy transfer from the fluid to the particles.
Since also escape depends on the shock hydrodynamics, this results in a \emph{universal} mechanism that returns extended power-law spectra in momentum.
At the zero-th order, these spectra are consistent with the astrophysical phenomenology of CRs and of the multi-wavelength emission from individual sources powered by non-relativistic shocks, such as novae, supernovae, heliospheric shocks, intra-cluster shocks, etc...
In \S\ref{sec:obs} we discuss where such predictions fail to explain observations and in \S\ref{sec:revised} how the DSA theory needs to be modified to resolve the discrepancy.

\subsection{The DSA Maximum Energy}
DSA is a very generic method for generating power-law spectra with roughly the expected slope, but it does not predict how many particles are injected into the acceleration mechanism \cite{blasi+05, caprioli+15}, and hence the overall normalization of $f$; 
also, it does not predict the maximum energy that can be achieved in a given source.
Note that a spectrum $\propto E^{-2}$ is mildly divergent (it has the same energy per decade), so a maximum energy $E_{max}$ is required in order to keep the total energy in CRs finite.
This energy limit may be due to the time needed to accelerate a CR up to $E_{max}$ or to the size of the accelerator, or even to energy losses particles may suffer.

The most relevant factor limiting the energy of cosmic hadrons is the acceleration time, which is related to the time a particle takes to diffuse back to the shock, i.e., the duration of a cycle, which increases with energy \cite{drury83}.
At some point, such an acceleration time becomes comparable with the age of the source (or the dynamical time-scale over which a shock must slow down), and a maximum energy is achieved. 

Given the spatial diffusion coefficient $D(E)$, the typical displacement from the shock scales in time as for a standard random walk $(\Delta x)^2\propto Dt$;
diffusion has to fight against a shock that ``chases'' the particle at a rate $\Delta x\propto \vsh t$ and hence, substituting  $\Delta x$, one derives a timescale for the residence time of the particle upstream of the shock 
\begin{equation}
    T_{acc}(E)\approx \frac{D(E)}{\vsh^2}.
\end{equation}
A more refined calculation that takes into account also the residence time in the downstream \cite{lagage+83a} returns
\begin{equation}
     T_{acc}(E)= \frac{3}{u_1-u_2}\left(\frac{D_1}{u_1}+\frac{D_2}{u_2}\right)\simeq \frac{6R}{R-1}\frac{D(E)}{\vsh^2},
\end{equation}
where in the last step we assumed $D\propto B^{-1}\propto u$, 
which is realized for Bohm diffusion since $D(E)=c/3 r_L(E)$, with $r_L(E)=\frac{E}{eB}$.

A similar constraint comes from requiring the diffusion length of an accelerating particle $\lambda\propto D(E)/\vsh$ not to exceed the source size $R_{\rm sh}\simeq \vsh t$.
This is analog to the Hillas criterion \cite{hillas84}, which states that the maximum energy achievable in a source corresponds to the potential drop of the motional electric field over the source extent. 
Note that for non-relativistic systems such an electric field is a factor of $\vsh/c$ smaller than the magnetic field, which means that the maximum allowed gyroradius is a factor $\vsh/c$ smaller than the system; 
for Bohm diffusion $\vsh/c$ is of the order of the ratio between the particle's gyroradius and diffusion length.

It is easy to show that, if particle diffusion occurred in SNRs at the same rate as in the interstellar medium \cite{evoli+21}, the maximum energy achievable by CRs would not be larger than a few GeV \cite{blasi05}.
Also considering a mean free path for pitch-angle scattering as small as the Larmor radius (Bohm diffusion), for the typical interstellar magnetic fields of 1--10 $\mu$G, it would be hard to account for energies as high as the knee \cite{lagage+83a,lagage+83b} and thus the SNR paradigm would fail.

\subsection{Magnetic field amplification in young SNRs}
A fundamental tile of the mosaic has been provided by the present generation of X-ray telescopes, such as Chandra and XMM--Newton.
Their high spatial resolution has revealed that young SNRs exhibit bright, narrow, rims, whose emission is due to synchrotron radiation by electrons with energies as high as 1--10 TeV \cite{morlino+10,ressler+14}.

These results are important for two reasons: first, they prove that SNRs accelerate electrons up to very high energies, and second, the measurement of the rim thickness provides a \emph{lower limit} to the magnetization of the post-shock medium.
Such non-thermal rims are not resolved by Chandra even in a close SNRs such as SN1006, which means that they are thinner than $\sim 0.01$pc.
For TeV electrons to lose energy on such small scales, the local magnetic fields must be of the order of a few hundreds $\mu$G, a factor 10--100 larger than typical interstellar ones, implying that some kind of \emph{magnetic field amplification} has to occur in young SNRs \cite{bamba+05,volk+05,parizot+06,caprioli+09a}.

This piece of information perfectly fits in the DSA paradigm: the diffusive flux of CR upstream of the shock has a net drift speed that is super-Alfv\'enic, which is expected to drive plasma instabilities that can change the topology and amplify any pre-existing magnetic field.
Such \emph{streaming instabilities} may lead to a rapid growth of different modes, either resonant \cite{kulsrud+68,skilling75a,bell78a} or non-resonant \cite{bell04,amato+09} with the Larmor radius of the relativistic particles, as discussed in other contributions in this volume.
%It is worth noticing that an efficient streaming instability (for which the magnetic field is expected to grow well beyond the background value) leads to a very fashionable scenario in which CRs provide by themselves a magnetic turbulence enhancing their diffusion.
Eventually, accounting for the CR-induced magnetic field amplification may allow SNRs to accelerate protons and heavier ions up to the knee \cite{blasi+07}, though the details depend on a complex chain in which non-linear wave-particle interactions make it hard to estimate the shape and the extent of the spectrum of CRs produced in different SN environments \cite{bell+13,cardillo+15, cristofari+21,cristofari+22}. 
Whether SNRs can be PeVatrons remains an outstanding question that seeks both theoretical and observational answers.

\section{The Need for a Non-linear Theory of DSA}\label{sec:challenges}
According to the SNR paradigm for the acceleration of Galactic CRs, a substantial fraction ($\gtrsim 10\%$) of the kinetic energy of the SN shocks should be converted into CRs.
As a consequence, at some point the validity of the \emph{test-particle} approach should break and  accelerated particles should not be considered passive spectators of the shock dynamics any longer.
In fact, as soon as the first quantitative calculations of the DSA efficiency were worked out, it became clear that pressure and energy in the shape of accelerated particles could no longer be neglected with respect to fluid ones, revealing the need for a \emph{non-linear theory of DSA} (NLDSA) in which particle acceleration and shock dynamics are self-consistently calculated.

The first attempts to carry out a study of NLDSA, often referred to as \emph{two-fluid} models, treated CRs as a fluid of relativistic particles \cite{axford+78,drury+81a,drury+81b}.
This approach is of great interest for pointing out the main effects of the non-linear CR backreaction on the shock.
Because of the CR pressure, which reaches its maximum at the shock (Eq.~\ref{eq:DSA}), the upstream develops a \emph{precursor} in which the fluid approaching the shock gradually slows down: 
the net result is to produce a weaker shock (now called the \emph{subshock}) and in turn a heating of the downstream plasma less efficient than in the test-particle case.
The presence of CRs  modifies the shock compression ratio in two ways: 
first, the contribution of relativistic particles to the total pressure makes the fluid more compressible, as if its adiabatic index $\to $ 4/3 and hence $R\to 7$; 
second, CRs escaping upstream because of the lack of confinement at large distances from the shock  act as a sink of energy, making the shock partially radiative. 
The net effect is that the far-upstream to downstream density ratio $\Rt$ becomes much larger than 4, while the subshock density ratio $\Rs$ is typically in the range 3--4.
Two-fluid approaches, though, cannot provide self-consistent information about the spectrum of the accelerated particles; 
this can be achieved via a \emph{kinetic} approach, in which CRs are described by means of a distribution function in both space and momentum.
Different ways of dealing with a kinetic analysis have been proposed in the literature, spanning from the semi-analytic models by \cite{malkov97,malkov+00,blasi02,blasi04,amato+05,amato+06,caprioli+10a,caprioli12}, to the Monte Carlo approach by \cite{ellison+85a,ellison+90,jones+91,ellison+95,ellison+96,vladimirov+06}, and to the numerical procedures by \cite{bell87,falle-giddings87,kang+97,kang+05,kang+06,kang+07,berezhko+99,kang13,kang+18}. 

The standard non-linear theory of DSA (e.g., \cite{jones+91,malkov+01} for reviews) suggests that CR spectra should deviate from power-laws, but confirms that the slope \emph{at a given momentum} should be determined by the effective compression ratio felt by such particles, which eventually depends on their diffusion length into the upstream.
High-$p$ particles that can probe the full precursor should feel a total compression ratio $\rt\gg 4$, and hence exhibit a spectrum that is \emph{flatter} than the standard $\propto p^{-4}$ prediction.
Since most of the pressure is carried by relativistic CRs, all particles with energy larger than a few GeV should thus exhibit spectra (much) flatter than the DSA prediction.
Conversely, non-relativistic CRs should be confined close to the subshock and have spectra slightly steeper than the standard prediction, since $\rs\lesssim 4$.
Accounting for the dynamical role of the magnetic turbulence produced by CRs themselves may limit such a compression to values less than $\sim 10$ \cite{caprioli+08, caprioli+09a}, but does not alter the theoretical expectation that shocks that are efficient in producing CRs should return concave spectra, invariably flatter than $p^{-4}$ at relativistic energies.

\subsection{The Challenging Observations}\label{sec:obs}
\begin{figure}[tb]
\centering
\includegraphics[trim=3px 3px 3px 3px, clip=true, width=.9\textwidth]{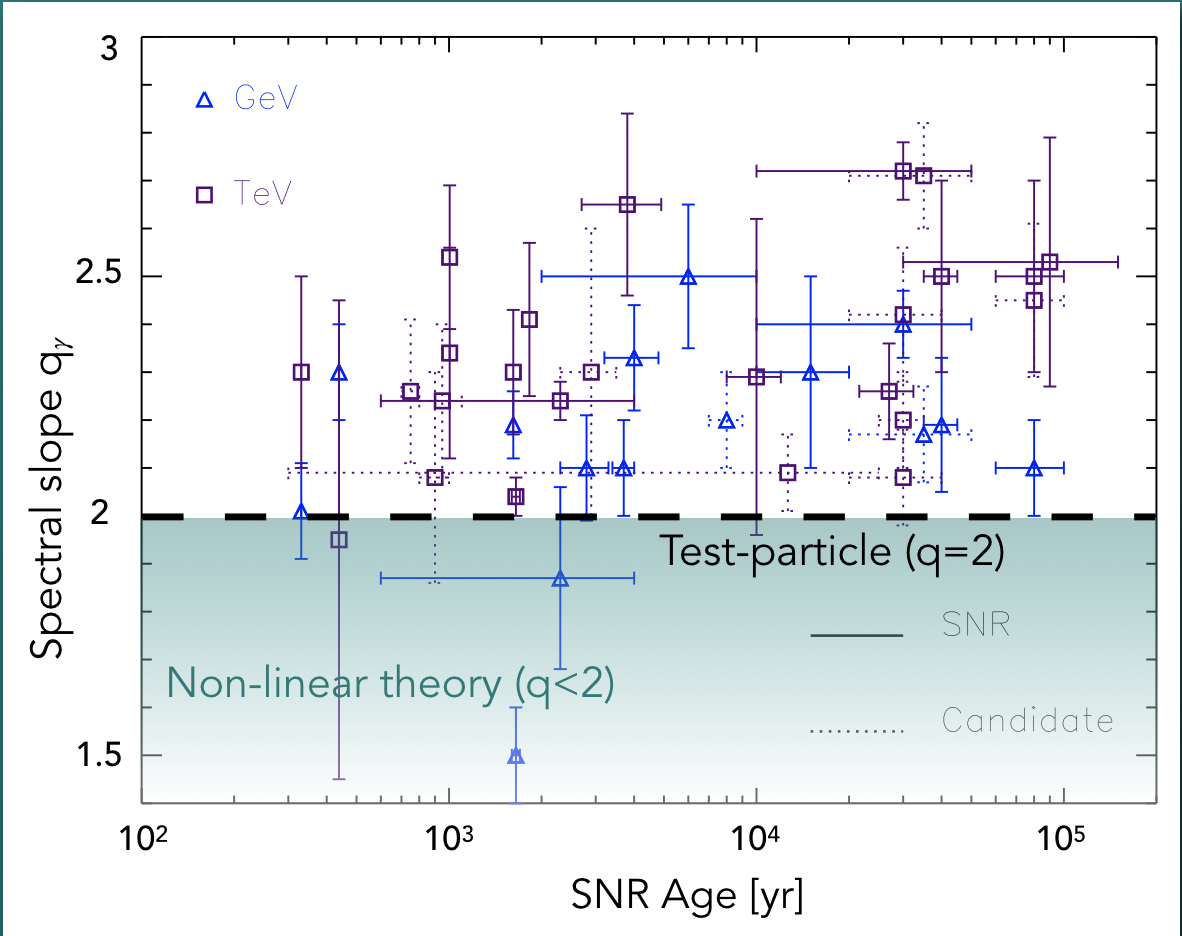}
\caption{Energy spectral indexes $q_{\gamma}$ of the $\gamma$-ray spectra observed in the GeV/TeV bands for SNRs of different ages.
Note that most of the SNRs show spectra steeper than the test-particle DSA prediction and much steeper than what expected when non-linear effects are included. 
Adapted from \cite{caprioli11,caprioli12}.}\label{fig:gamma}
\end{figure}

Since the NLDSA theory has been developed, all the attempts to find evidence of concave spectra in shock-powered astrophysical systems were at best inconclusive. 
Instead, more and more observations hinted to the fact that shock acceleration should produce particle spectra that are appreciably steeper than what predicted even by the test-particle DSA theory.
The main three pieces of evidence are summarized in the following.

{\bf $\gamma$-ray emission from SNRs.}
SNRs have been extensively observed by Cherenkov telescopes (HESS, MAGIC, VERITAS, and HAWC) in the TeV energy range and satellites (Fermi and AGILE) in the GeV band.
Very interestingly, the photon spectral index in most of the $\gamma$-ray bright SNRs is inferred to be appreciably larger than 2, typically in the range 2.2--3 \cite{caprioli11, caprioli12, acero+16short},  see Fig.~\ref{fig:gamma}.
$\gamma$-ray emission may be either  {\it leptonic} (relativistic bremsstrahlung and inverse-Compton scattering, IC) or {\it hadronic} ($\pi^0$ decay);
in the hadronic scenario, the $\gamma$-ray spectrum is parallel to the one of parent hadrons, while IC scattering produces harder photon spectra ($\propto E^{-1.5}$ for an $E^{-2}$ electron spectrum).
Therefore, away from the cut-off of the parent particle distribution, a steep spectrum represents a strong signature of hadron acceleration and suggests that the parent hadrons also have spectrum with $q>2$.
Remarkably, at GeV energies, where synchrotron cooling is not effective, protons and electrons are expected to show the same spectral index \cite{diesing+19,morlino+12}, which means that steep spectra are required even in a leptonic scenario.

{\bf  Radio-SNe.}
Very young SNRs (days to months old), observed in other galaxies in the radio and X-rays, also offer us clues that electrons are typically accelerated to relativistic energies with spectra as steep as $E^{-3}$ \cite{chevalier+06, soderberg+10, bell+11, margutti+19}; radiative losses does not seem to be important in this case, either.
Note that these so-called radio-SNe probe a different regime of shock acceleration than Galactic SNRs, the shock velocity still being quite large: $\vsh\gtrsim 10^4$ km s$^{-1}$ and even transrelativistic.

 {\bf Galactic CRs.}
Connected with the problem of accelerating CRs with power-law distributions is the problem of preserving such regular structures during the CR journey from sources to the Earth.
The CR Galactic residence time can be estimated thanks to radioactive clocks such as $^{10}$Be and to the   ratios of secondary to primary species  (e.g.,  B/C), which return the grammage traversed by primary CRs in the Galaxy.
%All of these measurements suggest that $\sim$10-GV CRs spend $\sim\,10^8$\,yr in the Galaxy before escaping, significantly longer that the ballistic propagation time.
If CRs are produced in the disk and diffusively escape at some distance $H$ ($\sim$\,a few kpc) in the halo, the Galactic residence time is $\tau_{\rm gal}(E)\approx H^2/D_{\rm gal}(E)$, where $D_{\rm gal}(E)$ is the diffusion coefficient that parametrizes CR transport in the Galaxy, assumed homogeneous and isotropic.
The energy dependence of primary/secondary ratios scales as $\tau_{\rm gal}\propto E^{-\delta}$ and is crucial for connecting the spectra injected at sources ($N_{\rm s}\propto E^{-\gamma}$) with those measured at Earth, ($\propto E^{-2.65}$ below the knee \cite{PAMELA14,ams16a}).
The equilibrium CR spectrum can in fact be written as $N_{\rm gal}(E)\propto N_{\rm s}(E) \mathcal{R_{\rm SN}}\tau_{\rm gal}(E)$, which imposes $\delta+\gamma\approx 2.65$.
Since the most recent AMS-02 data constrain $\delta\approx 0.33$  \cite{ams16a,evoli+21}, one finds $\gamma\approx 2.3-2.4$, quite steeper than the DSA prediction for strong shocks.
Note that the cumulative spectrum produced over the SNR history is typically $\sim 0.1$ steeper than the one at the beginning of the Sedov stage, i.e., $\gamma\simeq q_E+0.1$ \cite{caprioli+10a}.
Despite its simplicity, this diffusive model for CR transport is quite solid because it simultaneously accounts for the measured CR secondary/primary ratios, the diffuse Galactic synchrotron and $\gamma$-ray emission (see, e.g., \cite{blasi+11a,strong+98,evoli+19a,evoli+19b,evoli+21}), and even the observed anisotropy in the arrival directions of CRs \cite{blasi+11b}.

\section{Collisionless Shocks: Kinetic Simulations}  \label{sec:sims}
%\begin{figure}
%\centering
%\vspace{-0.7cm}
%\includegraphics[trim=0px 5px 0px 0px, clip=true, width=0.8\textwidth]{evo.pdf}
%\caption{\label{fig:p4fp}\footnotesize
%Time evolution of the post-shock ion energy spectrum for a parallel shock with $M=20$, showing a non-thermal tail that stems out of the thermal distribution for $E\gtrsim 2\esh$. 
%In the inset, the momentum spectrum is multiplied by $p^4$ to emphasize the agreement with DSA prediction at strong shocks.
%The downstream temperature is reduced by $\sim 20\%$ with respect to standard jump conditions, because of the energy going into accelerated particles \cite{DSA}.}
%\vspace{-0.3cm}
%\end{figure}

To address most of these questions it is necessary to model the non-linear interplay between energetic particles and the electromagnetic fields, which is very hard to tackle analytically. 
Astrophysical plasmas are typically collisionless, i.e., their dynamics is mediated by collective interactions rather than by binary collisions, and can be fruitfully modeled \emph{ab initio} by iteratively moving particles on a grid according to the Lorentz force and self-consistently adjusting the electromagnetic fields.  
Such particle-in-cells (PIC) simulations essentially solve the Vlasov equation by sampling the phase space with individual macro-particles and are particularly useful to account for spectra that may span several orders of magnitude in momentum, where standard Vlasov solvers may lose accuracy \cite{palmroth+18,juno20}.

While for understanding electron injection full PIC simulations are needed, the general dynamics of shocks should be sculpted by the accelerated ions;
therefore, one may revert to the \emph{hybrid} approach, in which electrons are considered as a massless neutralizing fluid \cite{lipatov02}, and still model shock formation, ion acceleration, and plasma instabilities self-consistently.
Hybrid simulations have been extensively used for heliospheric shocks\footnote{To give an idea, time and length scales accessible to hybrid simulations on modern supercomputers are comparable with the physical scales of the Earth's bow shock \cite{karimabadi+14}.} (e.g., \cite{giacalone+97,burgess+13}), and more recently even to stronger astrophysical shocks.
SNR shocks are characterized by large sonic and Alfv\'enic ($M_A\equiv \vsh/v_A$, with $v_A=B_0/\sqrt{4\pi m n}$ the Alfv\'en velocity) Mach numbers, which makes it computationally challenging to capture the diffusion length of accelerated ions $D/\vsh\approx v/\vsh r_L \gg  M_A c/\omega_p$ while resolving the ion skin depth $c/\omega_p$ ($\omega_p=\sqrt{4\pi n e^2/m}$ is the ion plasma frequency and $n, e,$ and $m$ the ion density, charge, and mass).
Also promising is the coupling of the hybrid technique with a MHD description of the background plasma \cite{bai+15,vanmarle+18}, though in this framework injection into DSA must be specified by hand.

Hybrid simulations have been used to perform a comprehensive analysis of ion acceleration at collisionless shocks as a function of the strength and topology of the pre-shock magnetic field, the nature of CR-driven instabilities, and the transport of energetic particles in the self-generated magnetic turbulence \cite{caprioli+14a,caprioli+14b,caprioli+14c}.  
Moreover, they have been used to unravel the processes that lead to the injection into DSA of protons \cite{caprioli+15}, ions with arbitrary mass/charge ratio \cite{caprioli+17}, and pre-existing CRs \cite{caprioli+18}. 

The progress in modeling non-relativistic shocks via first-principles simulations also features the first PIC simulations showing simultaneous acceleration of both ions and electrons \cite{park+15,kato13,xu+20,crumley+19,shalaby+22}, though it is still computationally challenging to go beyond 1D setups for long-term simulations.

\subsection{\label{sec:hybrid} Hybrid Simulations: Ion Acceleration}

Large 2D and 3D hybrid  simulations of non-relativistic shocks have been performed with the Newtonian code \emph{dHybrid} \cite{gargate+07} and its descendent \emph{dHybridR} \cite{haggerty+19a}, which allows accelerated particles to become relativistic.
The typical set up is outlined in \cite{caprioli+14a}: a supersonic/superalfv\'enic flow is smashed onto a reflective wall and the interaction between incoming and reflected flows produces a shock.
Lengths are measured in units of $c/\omega_p$, velocities normalized to the Alfv\'en speed $v_A$, and energies to $\esh\equiv m v_{\rm s}^2/2$, where $v_{\rm s}$ is the velocity of the upstream fluid in the downstream frame.
The shock strength is expressed by the Alfv\'enic Mach number $M_A$, assumed to be comparable with $M_{s}$ (both are indicated by $M$ if not otherwise specified).
The shock inclination is defined by the angle $\vartheta$ between the shock normal and the background magnetic field $\vec{B}_0$;
therefore, $\vartheta=0\deg~ (90\deg)$ for a parallel (perpendicular) shock.

\begin{figure}
\vspace{-0.0cm}
\centering
\includegraphics[trim=100px 0px 100px 0px, clip=true, width=.43\textwidth]{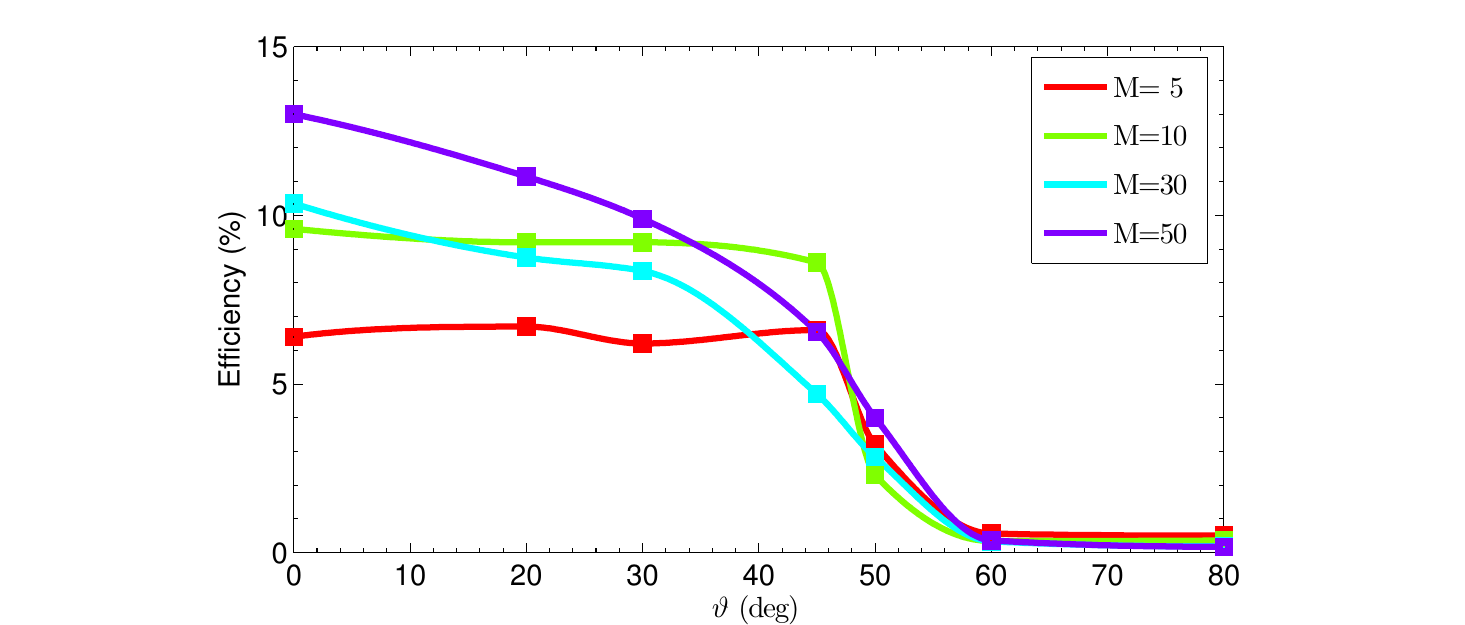}
\includegraphics[trim=110px 10px 90px 1350px, clip=true, width=.55\textwidth]{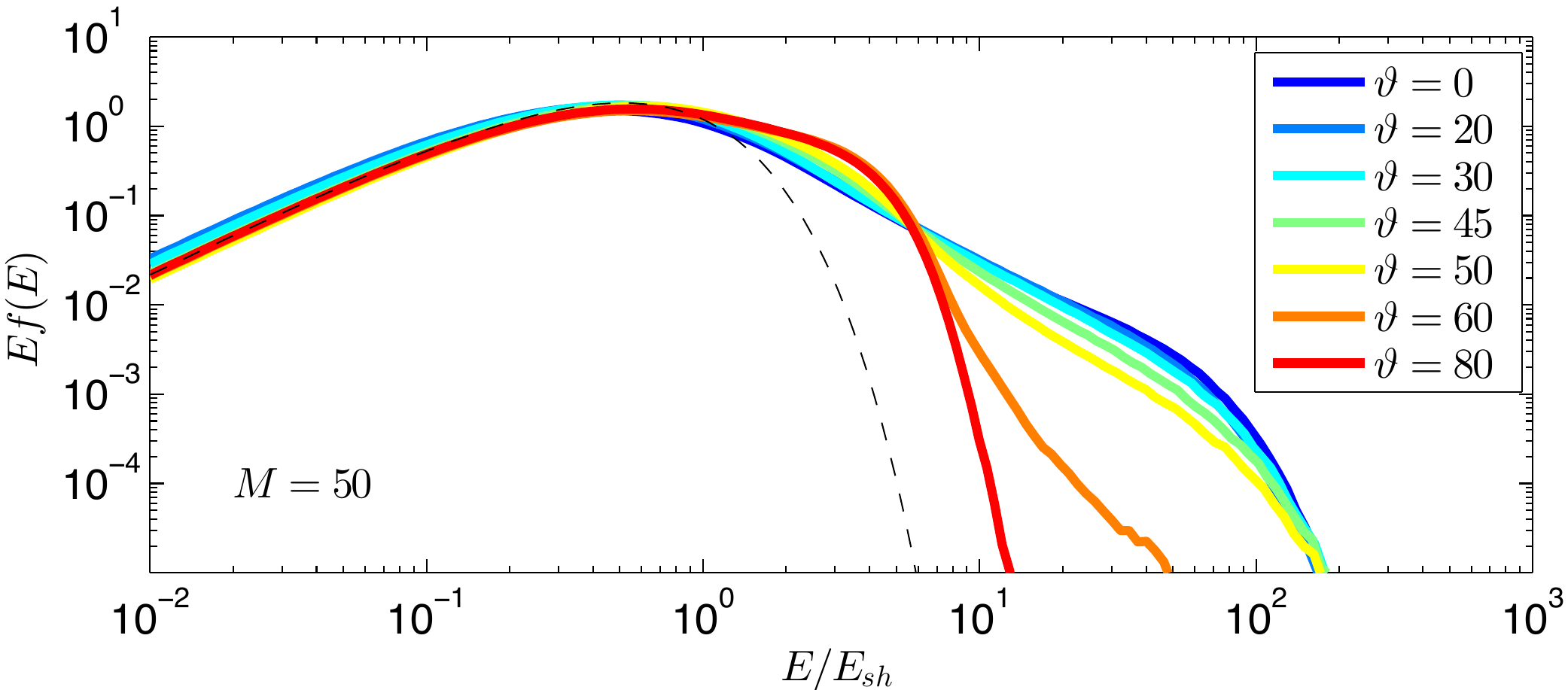}
\caption{\label{fig:eff}
\emph{Left Panel:} Fraction $\xcr$ of the downstream energy density in non-thermal particles as a function of shock inclinations and Mach numbers, $M$.
The largest acceleration efficiency ($\xcr\gtrsim 10\%$) is achieved for strong, parallel shocks, and drops for $\thbn\gtrsim 45\deg$, regardless of $M$.
\emph{Right Panel:} Post-shock particle spectra for $M=50$ and different shock obliquities, as in the legend.
The black dashed line represents the downstream Maxwellian.
Note how the non-thermal power-law tail develops only at low-inclination shocks  \cite{caprioli+14a}.
}
\vspace{-0.5cm}
\end{figure}
The kinetic simulations presented in refs.~\cite{caprioli+14a,caprioli+14b,caprioli+14c} have been able, for the first time, to demonstrate that DSA acceleration at non-relativistic strong shocks in general depends on the shock inclination and that can indeed be efficient.  

The left panel of Fig.~\ref{fig:eff} shows $\xcr$, i.e., the fraction of the bulk energy flow that goes into CRs as a function of shock strength and inclination. 
The acceleration efficiency can be as high as $\gtrsim 15\%$ at strong, quasi-parallel shocks, and drops for $\thbn\gtrsim 45\deg$, independently of the shock Mach number. 
The right panel of Fig.~\ref{fig:eff}, instead, shows the ion spectra for shocks with $M=50$ and different inclinations;
the DSA non-thermal tail vanishes for quasi-perpendicular shocks, where ions gain a factor of few in energy, at most.
Also 3D simulations show the same dependence of the acceleration efficiency on $\thbn$ \cite{caprioli+14a}.
Note that when CR acceleration efficiency is large, the post-shock temperature is accordingly reduced with respect to Rankine--Hugoniot jump conditions;
such a modification is a manifestation of the back-reaction of efficient CR acceleration, as predicted by most models of non-linear DSA (more on this in \S\ref{sec:modified}).

\begin{figure}[t]
\centering
\includegraphics[trim=5px 5px 10px 1px, clip=true, width=0.8\textwidth]{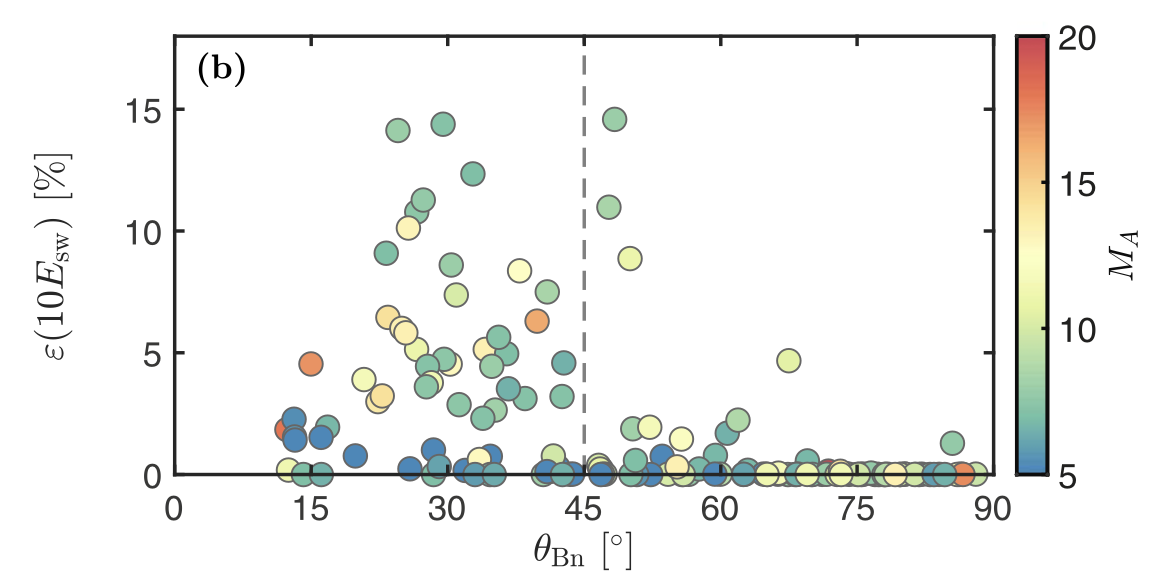}
\caption{\label{fig:theta}
MMS observations downstream of the Earth's bow shock. $\varepsilon$ is the fraction of the energy flux in ions with more than 10 times the solar wind kinetic energy \cite{johlander+21}.
Efficiency is highest at quasi-parallel ($\theta\lesssim45\deg$) shocks, as predicted by simulations (see Fig.~\ref{fig:eff}).}
\end{figure}

\subsection{\label{sec:MFA} Magnetic Field Amplification}
Since the initial formulation of the DSA theory, particle acceleration has been predicted to be associated with plasma instabilities \cite{bell78a}, in particular with the generation of magnetic turbulence at scales comparable to the gyroradii of the accelerated particles (\emph{resonant streaming instability}, see \cite{skilling75a,bell78a}). 
Then, Bell pointed out that non-resonant, short-wavelength modes may grow faster than resonant ones (\emph{non-resonant hybrid instability} \cite{bell04}). 

\begin{figure}
\begin{center}
%\vspace{-0.7cm}
\includegraphics[trim=45px 75px 60px 55px, clip=true, width=.85\textwidth]{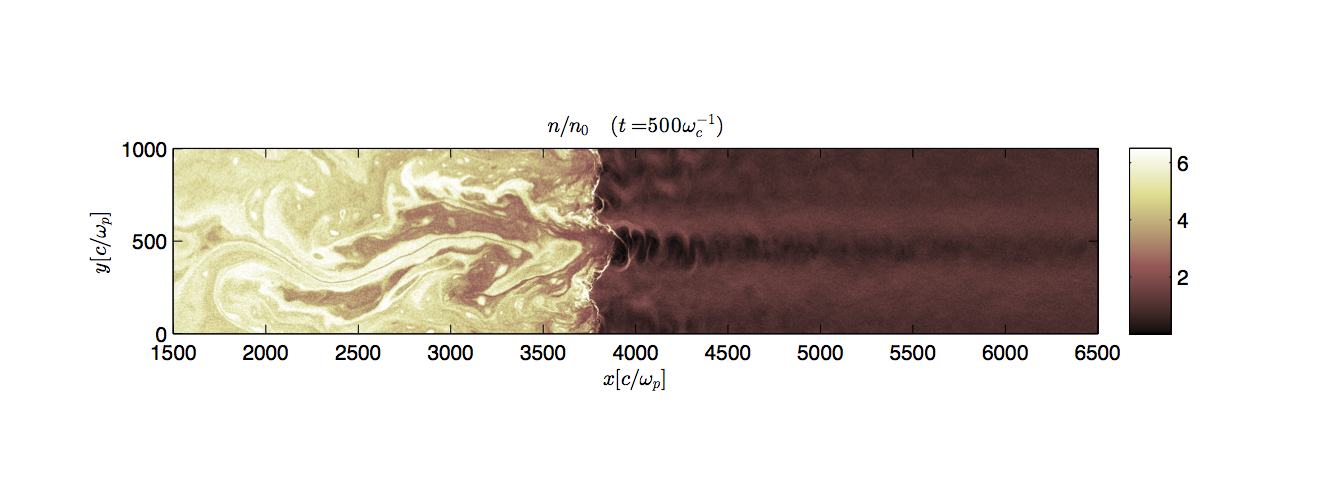}
\includegraphics[trim=35px 75px 60px 55px, clip=true, width=.85\textwidth]{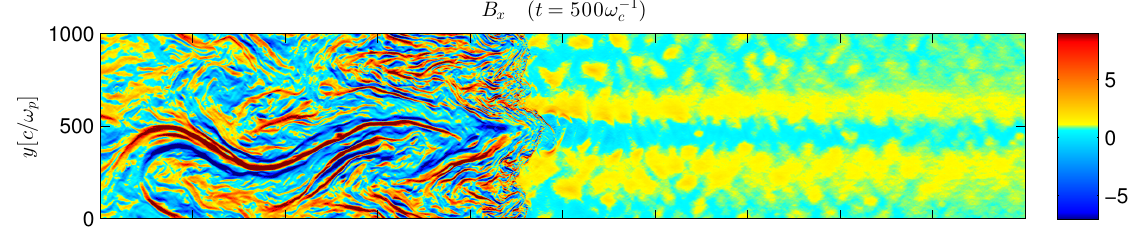}
\includegraphics[trim=45px 75px 60px 55px, clip=true, width=.85\textwidth]{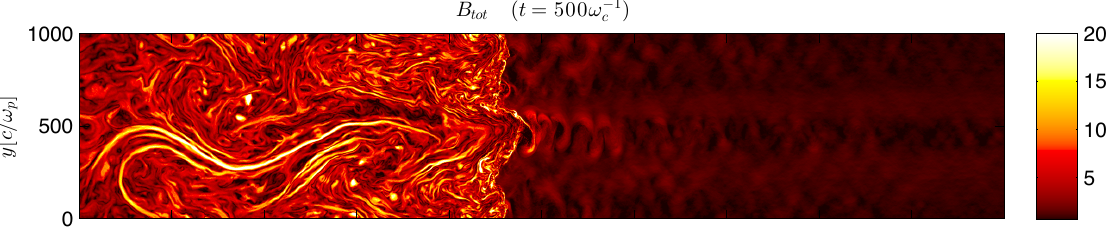}
\includegraphics[trim=35px 75px 60px 55px, clip=true, width=.85\textwidth]{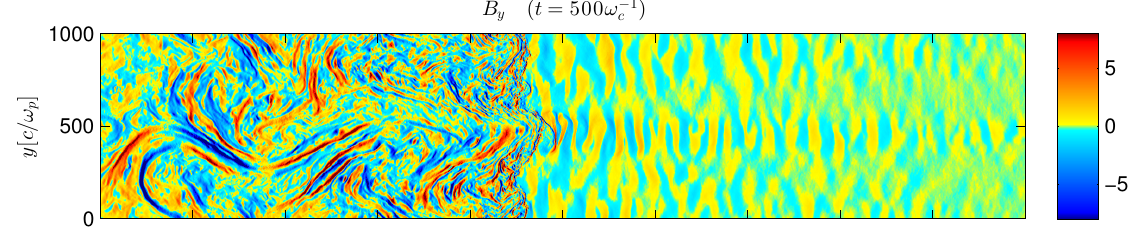}
\includegraphics[trim=35px 45px 50px 45px, clip=true, width=.85\textwidth]{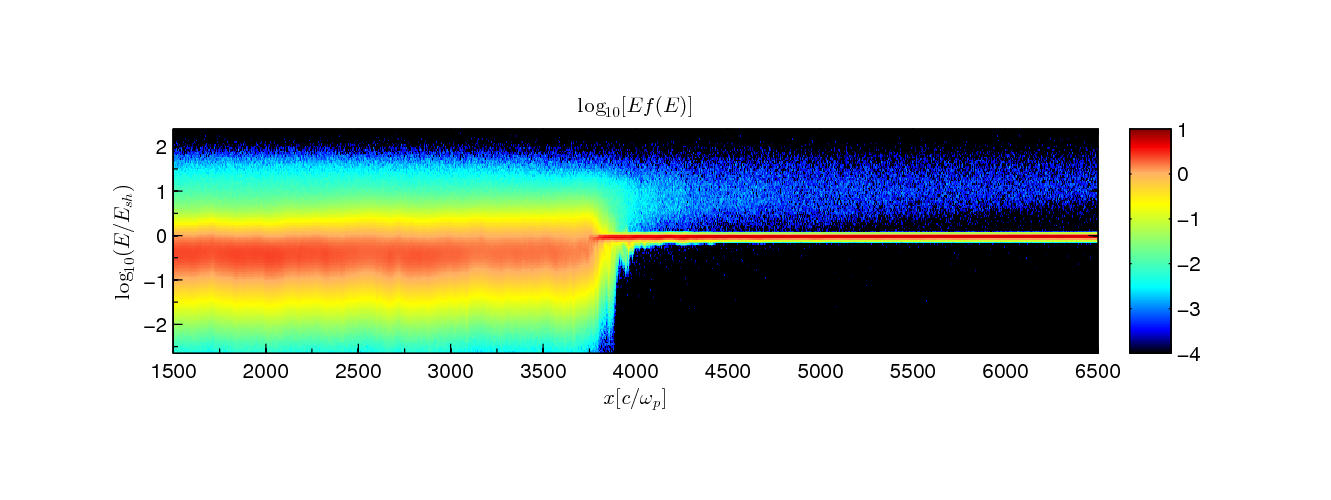} 
\includegraphics[trim=35px 45px 60px 55px, clip=true, width=.85\textwidth]{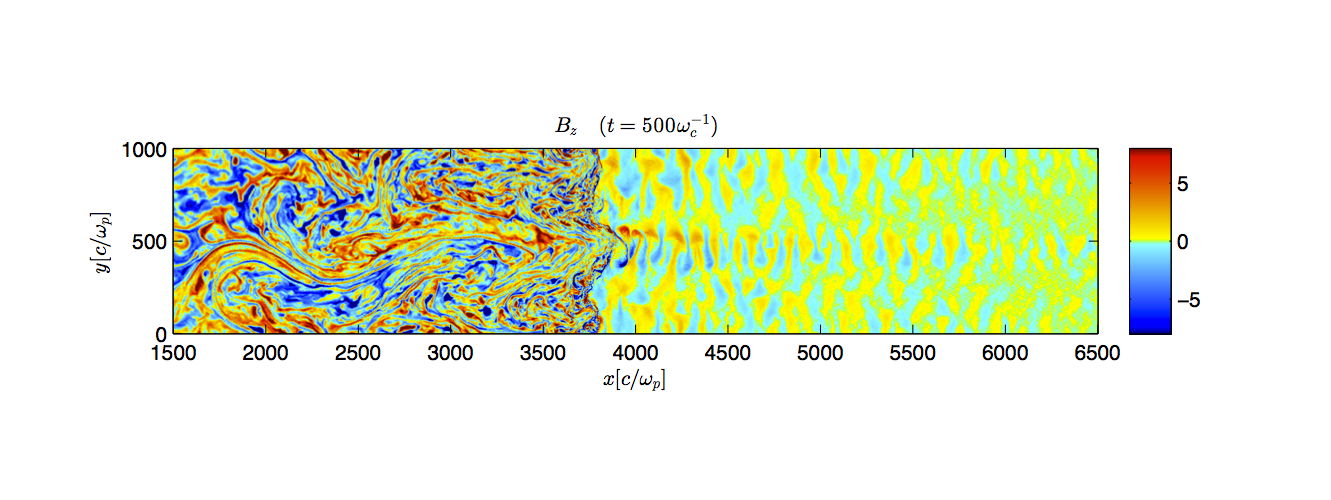}
\end{center}
\vspace{-0.5cm}
\caption{\label{fig:hybrid}
Output of a 2D hybrid simulation of a parallel shock with $M=30$. 
\emph{Left panels}: density; total magnetic field strength; ion $x-p_x$ phase space. \emph{Right panels}: components of the magnetic field ($B_x$, $B_y$, $B_z$). 
Such a rich structure is entirely generated by instabilities driven by accelerated ions diffusing in the upstream (to the right in the figures) \cite{caprioli+13}.}
\vspace{-0.5cm}
\end{figure}

Fig.~\ref{fig:hybrid} shows the structure of a parallel shock with $M=30$, with the upstream (downstream) to the right (left).
In the shock precursor, a cloud of non-thermal particles drives a current able to amplify the initial magnetic field by a factor of a few, also leading to the formation of underdense cavities filled with energetic particles and surrounded by dense filaments with strong magnetic fields.
The typical size of the cavities is comparable with the gyroradius of the highest-energy particles (a few hundred ion skin depths for the simulation in Fig.~\ref{fig:hybrid}).
Note that, when the streaming instability enters its non-linear stage, filamentary modes are expected to grow, too (e.g., \cite{reville+12,caprioli+13,bykov+13}). 

The propagation of the shock through such an inhomogeneous medium leads to the formation of turbulent structures (via the Richtmyer--Meshkov instability), in which magnetic fields are stirred, stretched, and further amplified.
In this case, amplification via turbulent dynamo may become important even in the absence of large pre-existing density fluctuations.

Magnetic field generation depends on the presence of diffuse ions, hence it is more prominent at quasi-parallel shocks.
Simulations show that the maximum amplification achieved in the precursor scales as $\delta B/B_0\propto\sqrt{M}$ and ranges from factors of a few for $M\lesssim 5$ to factors of $\gtrsim 7$ for $M\gtrsim 50$ (see  figure 5 in \cite{caprioli+14b}).
For $M\gtrsim 20$ the non-resonant instability grows significantly faster than the resonant one \cite{riquelme+09,gargate+10}, exciting distinctive right-handed modes with wavelength much smaller than the gyroradius $r^*_L$ of the CRs driving the current.
Then, in the non-linear stage, an inverse cascade in $k-$space progressively channels magnetic energy into modes with increasingly small wavenumber $k$. 
The non-resonant instability eventually saturates when the maximally-growing mode is $k_{\max} \approx 1/r^*_L$, which effectively scatters the current ions \cite{zacharegkas+22}.
This is the very reason why the resonant instability saturates already when $\delta B/B_0\sim1$ \cite{mckenzie+82} while the non-resonant one can grow up to non-linear levels before the driving current is disrupted.
For $M\lesssim 10$, $\delta B/B_0\lesssim1$ and both wave polarizations are observed, consistently with the prediction of quasi-linear theory \cite{amato+09}.
The reader can refer to \cite{caprioli+14b} for a more detailed discussion of the wave spectra and the saturation of the two instabilities.

\subsection{\label{sec:diff}Particle Diffusion}
%Magnetic field amplification is required for fast CR energization.
CRs are scattered in pitch angle by waves with resonant wavenumbers $k(p)\sim 1/r_L(p)$; in the regime of small deflections this process can be described by a diffusion coefficient. 
The most popular choice is to assume the Bohm limit, which is obtained (in the quasi-linear limit $\delta B/B_0\lesssim 1$) for an Alfv\'enic turbulence generated via resonant streaming instability by a CR distribution $\propto p^{-4}$ \cite{bell78a}.
Bohm diffusion is often heuristically extrapolated into the regime of strong field amplification, but such a prescription used to lack a solid justification.  

\begin{figure}\centering
%\vspace{-0.7cm}
\includegraphics[trim=0px 50px 0px 260px, clip=true, width=.485\textwidth]{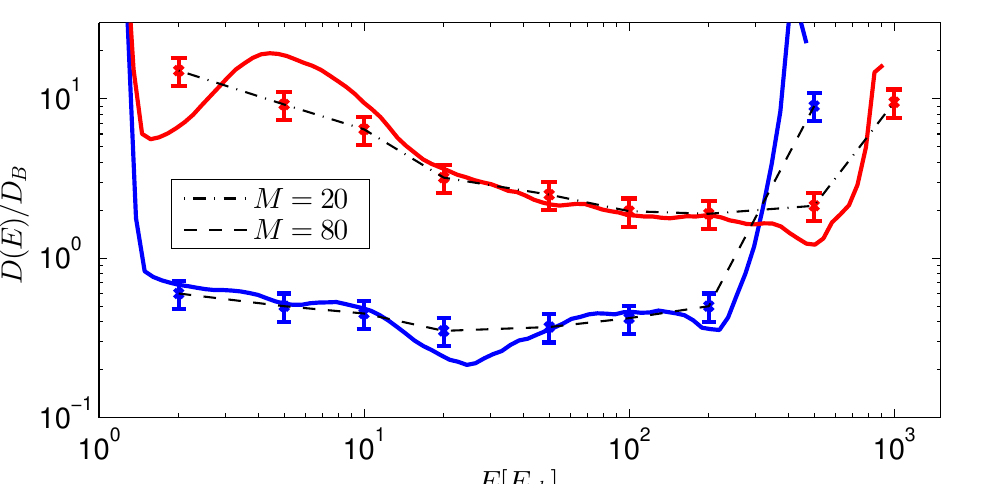}
\includegraphics[trim=30px 0px 40px 15px, clip=true, width=.485\textwidth]{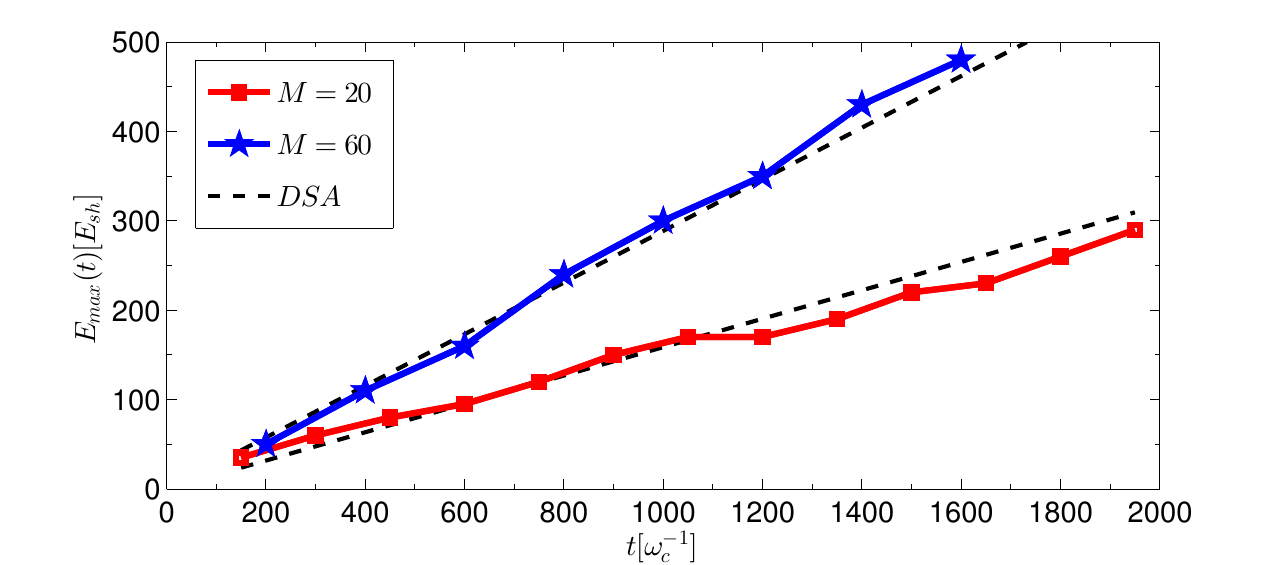}
\caption{\label{fig:cd}
{\emph{Left panel}: Diffusion coefficient, normalized to Bohm, immediately in front of the shock for  shocks with $M=20, 80$.
\emph{Right panel}: Time evolution of the maximum ion energy for parallel shocks with $M=20,60$, compared with the DSA prediction  (dashed lines) \cite{caprioli+14c}.}
}
\vspace{-0.5cm}
\end{figure}

Global hybrid simulations allow reconstructing CR diffusion in different regions of the shock, either by using an analytical procedure based on the extent of the CR distribution in the upstream or by tracking individual particles \cite{caprioli+14c}.
The two methods return consistent results, as shown in the left panel of Fig.~\ref{fig:cd} (see \cite{caprioli+14c} for more details).
When magnetic field amplification occurs in the quasi-linear regime ($M\lesssim20$), particle scattering is well described by the diffusion coefficient self-generated via resonant instability \cite{bell78a,amato+06}, where the scattering rate depends on the magnetic power in resonant waves.
For stronger shocks, instead, $D(E)$ is roughly proportional to the Bohm coefficient and its \emph{overall} normalization depends on the level of magnetic field amplification $\delta B/B_0\gtrsim 1$ (see also \cite{reville+13}).
Such a scaling is determined by the fact that far upstream the spectrum of the excited magnetic turbulence (see \cite{caprioli+14b}, Figs.\ 6 and 7) peaks at relatively large wavelengths, comparable with the gyroradius of the highest-energy ions.  

The effective scattering rate is also imprinted in the time evolution of $\Em$. 
The right panel of Fig.~\ref{fig:cd} shows such an evolution, which is linear with time with a slope inversely proportional to the measured diffusion coefficient (dashed lines), as expected for DSA (e.g., \cite{drury83,lagage+83a,blasi+07}).

%However, such flat spectra are never observed in SNRs \cite{caprioli11,caprioli12} or in radio SNe \cite{chevalier+06}, and would be inconsistent with the observed fluxes of Galactic CRs when propagation in the Milky Way is accounted for \cite{blasi+11a,evoli+19b}.
%The reader is referred to \cite{caprioli15p,caprioli+19p,caprioli+20} and references therein for more details about the tension between DSA theory and observations, and for the possible solutions that have been suggested in the literature. 

%In this work we present the results obtained with kinetic simulations of non-relativistic shocks \cite{haggerty+20,caprioli+20} that lay the groundwork for a theory of efficient DSA in which CR-modified shocks show both enhanced compression and steeper spectra with respect to the test-particle predictions.

\subsection{Oblique Shocks and the Importance of 3D}\label{sec:3D}
\begin{figure}[t]
\begin{center}
    \includegraphics[width=0.75\textwidth]{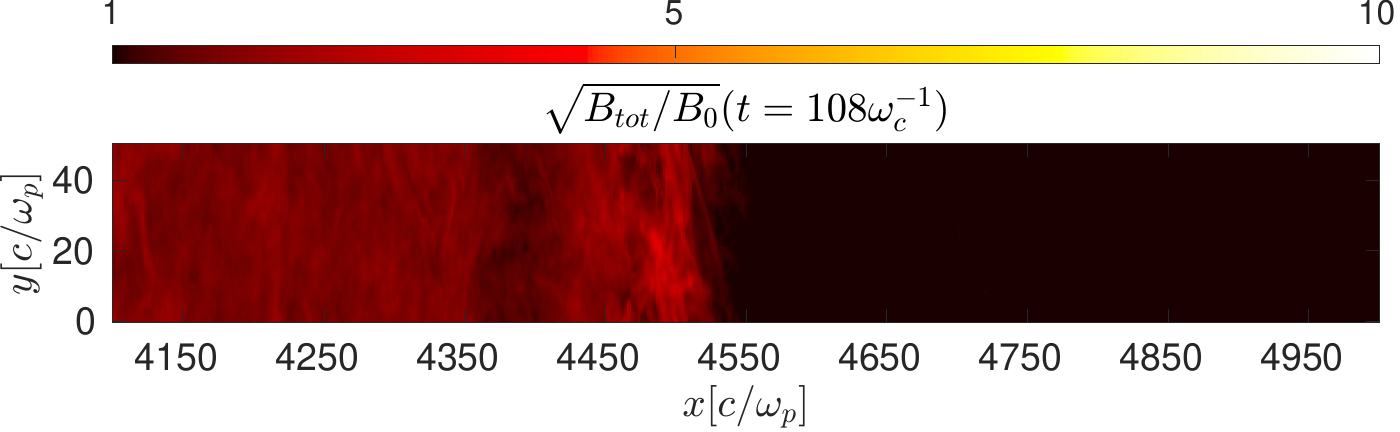}
    \\
    \includegraphics[width=0.75\textwidth]{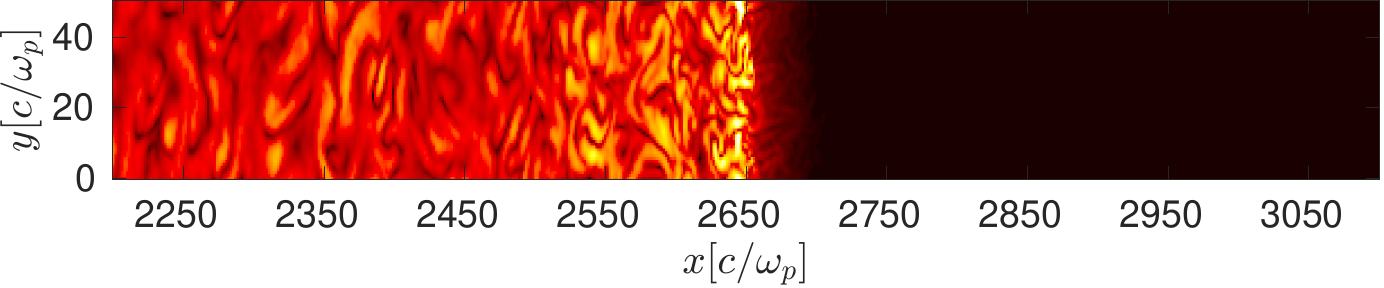}
    \\
    \includegraphics[width=0.75\textwidth]{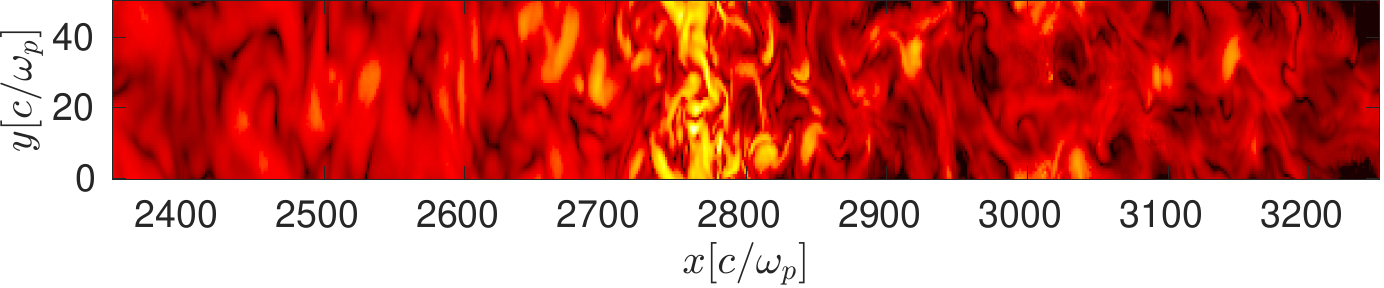}
    \includegraphics[width=0.75\textwidth]{B_evolution.pdf}
    \end{center}
    \caption{From the top: Total magnetic field for the out-of-plane (2D$z$), in-plane (2D$y$), and 3D setups. Bottom panel: evolution of $B_{\rm tot}$ in 3D. In all cases $M$=100 and $t=108 \omega_c^{-1}$.
    Note how 2D$z$ captures the ion-Weibel instability at the shock, and hence the downstream generation of turbulence, but 3D is needed to reveal the upstream $B$-field amplification due to back-streaming (injected) energetic ions.} 
    \label{Fig:B_field}
\end{figure}

\begin{figure}
\begin{center}
    \includegraphics[width=0.75\textwidth]{"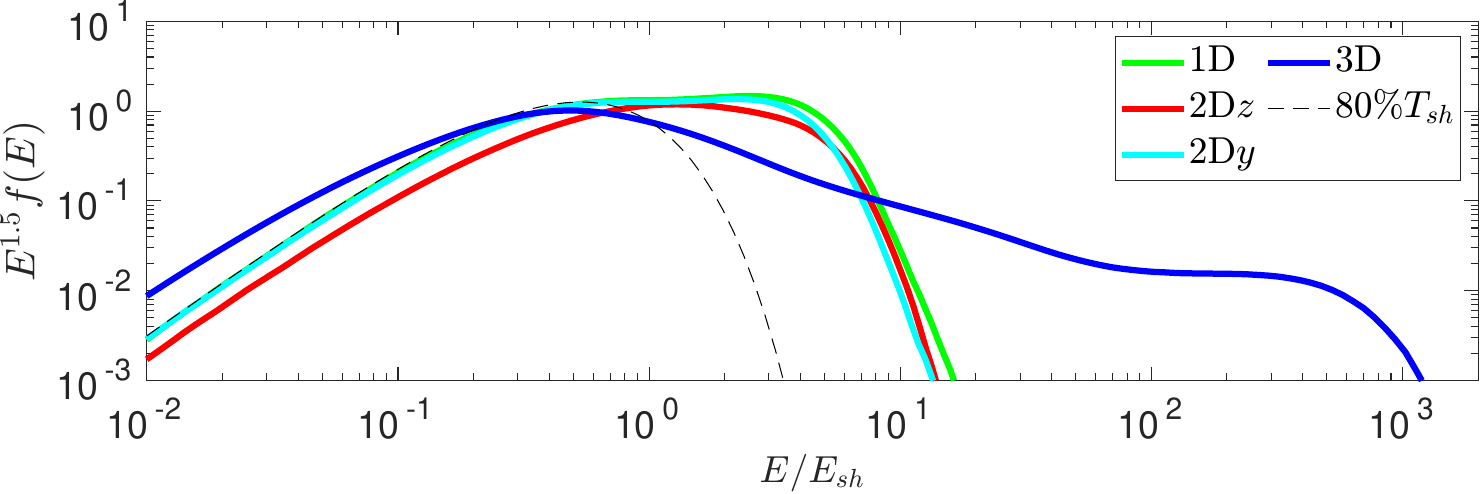"}
    \includegraphics[width=0.75\textwidth]{"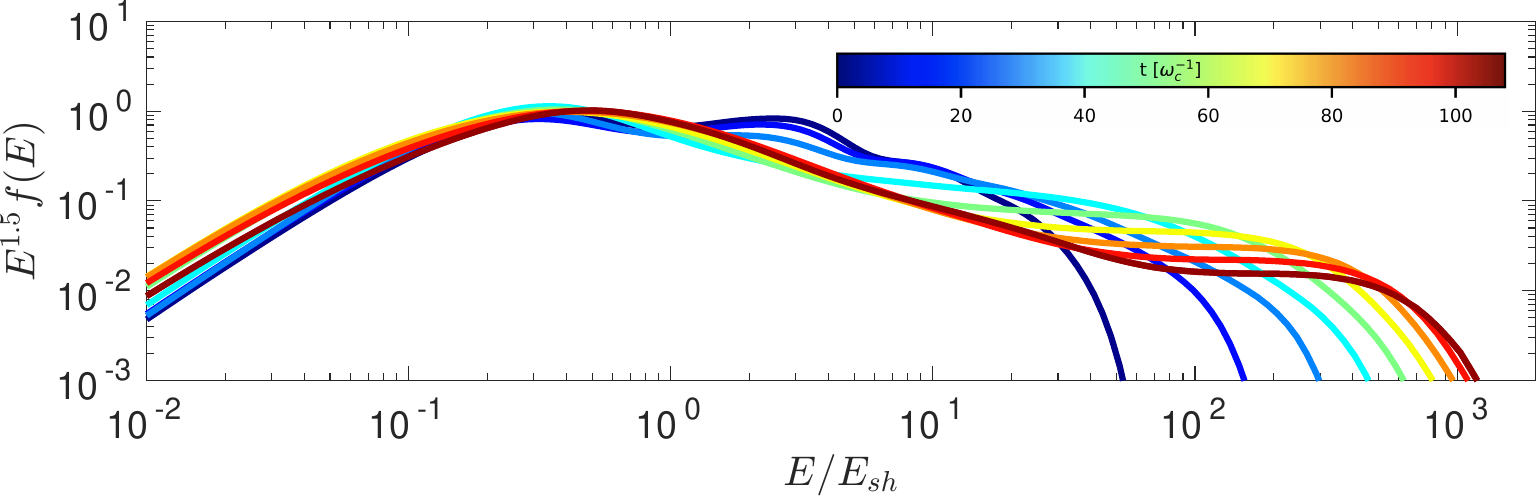"}
    \includegraphics[width=0.75\textwidth]{"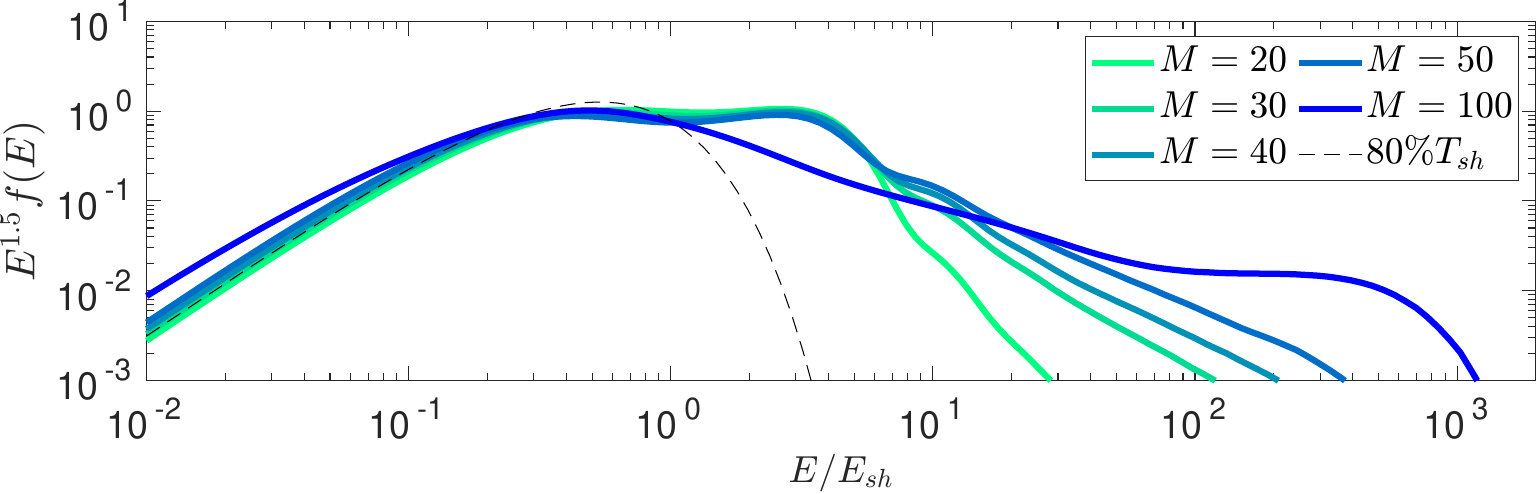"}
    \caption{Top panel: downstream ion spectrum  at $t=108 \omci$ for $M=100$ and different setups.
    A non-thermal tail above $10\esh$ is present only in 3D, while  1D and 2D show similar truncated spectra. 
    Middle panel: time evolution of the ion spectrum in 3D. 
    Bottom panel: spectrum dependence on $M$; the dashed line indicates a Maxwellian with temperature $\sim 80\%$ of the one expected for a purely gaseous shock.
    3D is necessary to capture the cross-field diffusion that leads to ion injection at relatively-large $M$ numbers (low magnetization). } 
    \label{Fig:Spectrum_example}
    \end{center}
\end{figure}

The hybrid campaign detailed in \cite{caprioli+14a,caprioli+14b,caprioli+14c, haggerty+20,caprioli+20} focused on 2D simulations, plus some 3D runs with $M=6$ \cite{caprioli+14a}. 
One natural question that arises is: \emph{Should the direction of the upstream magnetic field matter if the shock magnetization becomes smaller and smaller?} (i.e., if $M$ is sufficiently large?).
Orusa and Caprioli \cite{orusa+23} performed a study of cases with  $M\gtrsim 10$ and found, for the first time in kinetic simulations, that  in 3D a non-thermal tail develops \emph{spontaneously}, i.e., \emph{without pre-existing seeds or turbulence} \cite{giacalone05,lembege+04, caprioli+18}; ions perform multiple shock drift acceleration cycles before either being advected downstream or escaping upstream.

Oblique and quasi-perpendicular shocks are known to potentially be \emph{fast accelerators} \cite{jokipii87,giacalone+94} but are also known to be less effective than quasi-parallel shocks in injecting thermal particles \cite{schwartz+83,caprioli+15,caprioli+14a}.
On the other hand, oblique heliospheric shocks are often associated with particle acceleration (mostly electrons, less frequently ions) \cite{zank+06,zank+07}.
While PIC simulations easily reproduce populations of back-streaming electrons \cite{amano+07,xu+20,guo+14a,guo+14b,morris+23,ha+21,ha+22,bohdan+19a,bohdan+21,amano+22,matsumoto+15}, 3D simulations are needed to study of injection and acceleration of thermal ions.
This likely explains why acceleration efficiency in 2D hybrid simulations cuts off quite abruptly for $\thbn\gtrsim 45\deg$, different from the shallower trend in MMS events at the Earth bow shock (Fig.~\ref{fig:theta}).
 
As pointed out in \cite{jokipii+93,jones+98,giacalone+00}, cross-field diffusion plays a crucial role in the return of ions from downstream, and is not properly captured if not in 3D (see fig. \ref{Fig:B_field}).
Jones et al. \cite{jones+98} demonstrated that charged particles in an arbitrary electromagnetic field with at least one ignorable spatial coordinate remain forever tied to a $B$-field line.
Since in 2D field lines are effectively transverse ``sheets", ion diffusion along the shock normal is inhibited;
in 3D, instead, field lines can twist and intertwine, and ions can diffuse cross-field, which effectively prevents them from being rapidly swept downstream.
Tracking reveals that in 2D ions are advected downstream after a couple of gyrations, while in 3D they diffuse back several times, gaining energy at each SDA cycle. 
To some extent, this acceleration mechanism is similar to that proposed by \cite{kamijima+20}, who argued that the extreme case of a perpendicular shock where Bohm diffusion were realized downstream would be a rapid accelerator;
our self-consistent simulations show that the process may occur only for large $M$ and may be intrinsically limited when $\thbn<90\deg$.

Particles initially gain energy through shock drift acceleration (SDA), tapping into the motional electric fields $E=- \frac{v}{c} \times B$ during their gyrations around the shock \cite{chen+75,ball+01}.
Acceleration then briefly transitions to DSA (where $\Em\propto t$) before reaching a limit energy $\Em^*$, beyond which particles escape upstream. 

Acceleration efficiency and spectral slope quite strongly depend on the shock Mach number $M$: while for $M\lesssim 20$ efficiency is only a few percent and spectra are very steep, for $M\gtrsim 50$ efficiency can exceed 10--20$\%$ and spectra converge to the DSA ones, as flat as $p^{-4}$ in momentum (Fig.~\ref{Fig:Spectrum_example});
also the level of magnetic field amplification and the maximum energy limit increase with $M$.

The biggest questions that remain open are whether oblique/quasi-perpendicular shocks can efficiently drive plasma instabilities strong enough to self-sustain DSA up to energies significantly larger than $\Em^*$, and whether the same acceleration process is viable for electrons, too.
Both questions require different numerical approaches that are  capable of either capturing the longer-term evolution of the system or the physics of electron injection.

\section{Hybrid Simulations Reveal Cosmic-ray--modified Shocks}\label{sec:modified}

The \dHybridR~code, which allows accelerated particles to become relativistic \cite{haggerty+19a}, has been recently used to investigate the long-term evolution of non-relativistic shocks.
The reader can refer to \cite{haggerty+20,caprioli+20} for more information about the setup; 
one important ingredient in this campaign is that the authors assumed the fluid electrons to be adiabatic, rather than prescribing an effective polytropic index $\gamma_e$ aimed to enforce electron/ion equipartition downstream (see appendix of \cite{caprioli+18}).
The latter choice, in fact, requires to fix the compression ratio a priori: if one guesses $r=4$, the electron equation of state becomes very stiff ($\gamma_e\sim 3$) and prevents any shock modification, enforcing $r\sim 4$ \cite{caprioli+14a}. 
Instead, using $\gamma_e\sim 5/3$, or iteratively setting $\gamma_e$ until equipartition is self-consistently achieved even for $r>\sim 4$, yields consistent results.

Let us consider as benchmark a strong shock, with both sonic and Alfv\'enic Mach number $M=20$, propagating along a background magnetic field (parallel shock).
In this case, about 10\% of the shock kinetic energy is converted into accelerated particles \cite{caprioli+14a}.

\subsection{CR-induced Precursor and Postcursor}

Such a benchmark run confirmed the prediction that, when DSA is efficient, the shock develops an upstream precursor, in which the incoming flow is slowed down and compressed under the effect of the CR pressure (see figure 2 of \cite{haggerty+20}).
What was unexpected is that the shock also develops a \emph{postcursor}, i.e., a region behind the shock where the dynamics is modified by the presence of CRs and self-generated magnetic perturbations. 
More precisely, \cite{haggerty+20} attest to the presence of an extended region in which magnetic structures drift at a finite speed \emph{towards downstream infinity} with respect to the thermal gas.
A Fourier analysis (figure 7 of \cite{haggerty+20}) shows that the phase speed of the magnetic fluctuations is comparable to the local Alfv\'en speed, both upstream and downstream;
as a result, CRs ---which tend to become isotropic in the wave frame---  also have a comparable net drift with respect to the background plasma (see \cite{haggerty+20}, fig. 5). 

The development of the postcursor implies that energy/pressure in CRs and magnetic fields are advected away from the shock at a faster rate that in a gaseous shock, which has has two crucial effects: 
1) it makes the shock behave as partially radiative, enhancing its compression;
2) it makes the CR spectrum steeper, enhancing the rate at which particles leave the acceleration region.

\subsection{Enhanced Shock Compression Ratio}
\begin{figure}[t]
\includegraphics[width=.45\textwidth,clip=true,trim= 1 0 0 0]{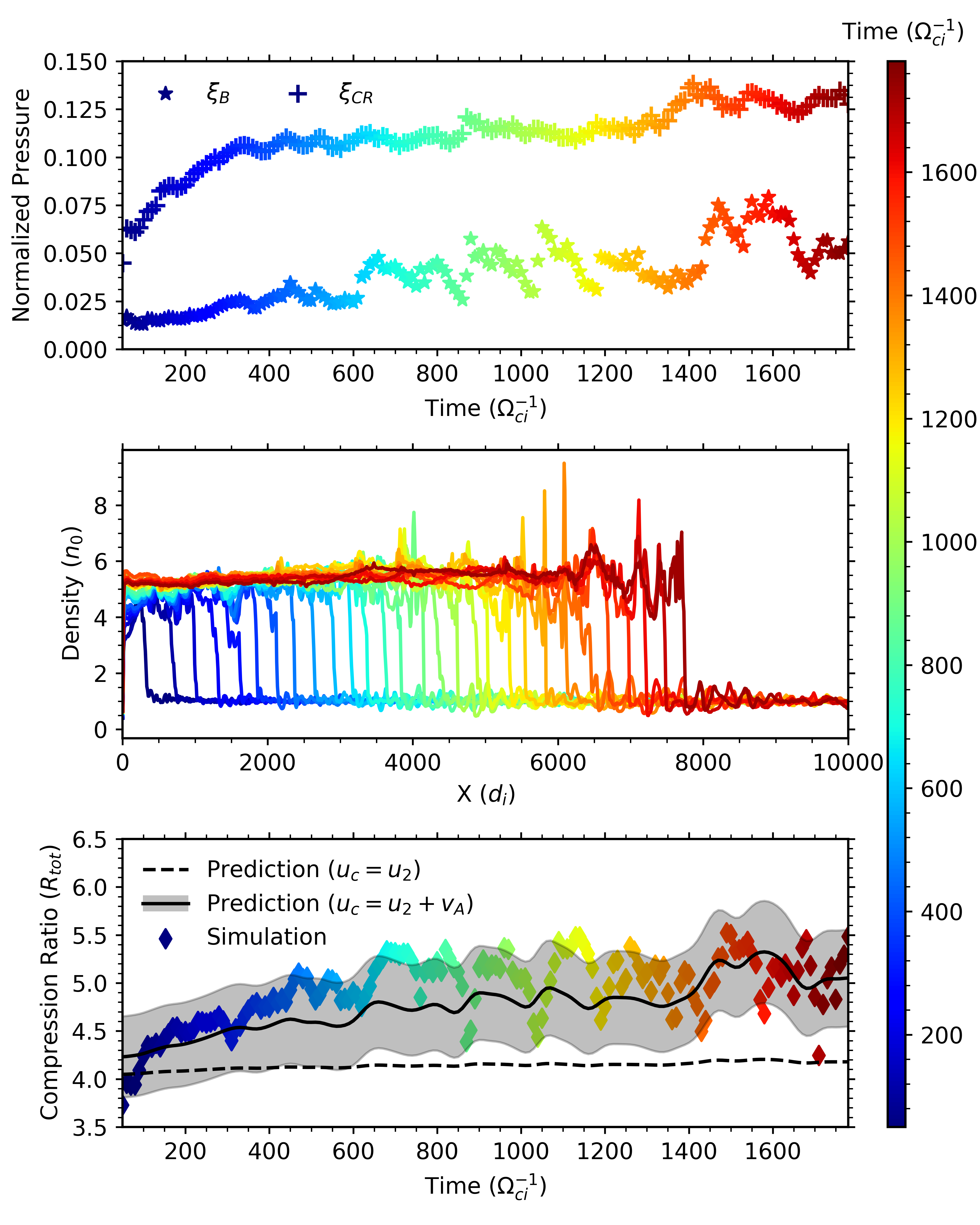}
\includegraphics[width=0.55\textwidth, clip=true,trim= 1 0 50 16]{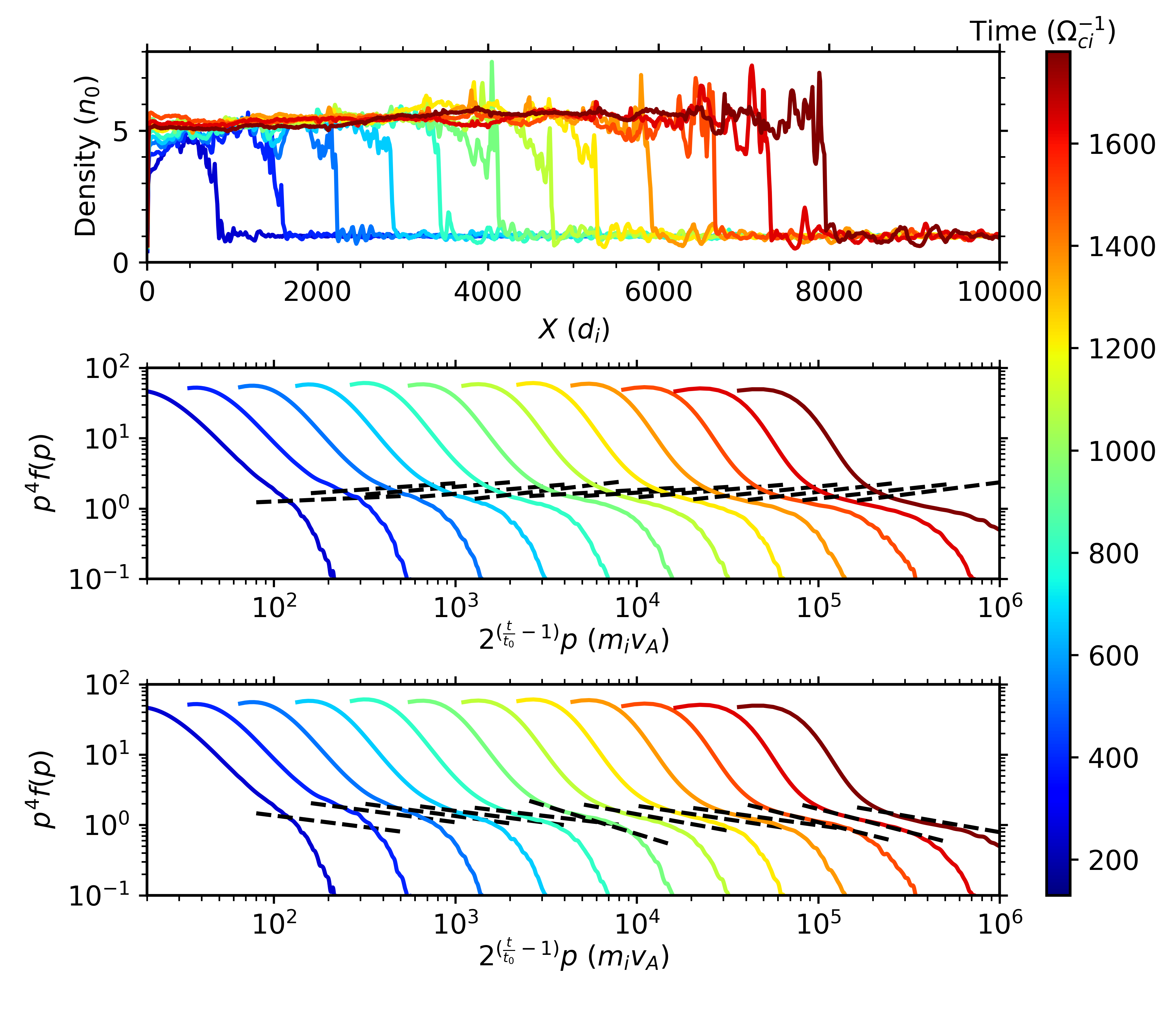}
\caption{
Time evolution (color coded) of physical quantities that show the CR-induced modification.
Left panels: normalized magnetic and CR pressures in the postcursor, $\xi_B$ and $\xi_c$  (top), density profile (middle), and total compression ratio, $\rt$ (bottom).
The CR pressure quickly converges to $\xi_c\approx 10\%$, while $\xi_B$ saturates around 6\%;
at the same time, the compression ratio departs from the test-particle value of $\sim 4$ and becomes $\rt>\sim 5.5$.
The prediction including the postcursor effect \cite{haggerty+20} is shown in gray.
Right panels: density profile (top) and post-shock particle spectra compared with standard (Equation \ref{eq:DSA} with $r\to\rt$, middle) and revised prediction (Equation \ref{eq:qtilde}, bottom).
Despite the fluid shock compression ratio $\rt\to 6$, the CR non-thermal tail is significantly steeper than $p^{-4}$.}\label{fig:R}
\vspace{-0.5cm}
\end{figure}

The left panels of Fig.~\ref{fig:R} illustrate the hydrodynamical modification induced by the postcursor. 
The normalized pressure in CRs and magnetic fields, $\xi_c$ and $\xi_B$, are plotted as a function of time in the first panel (crosses and stars, respectively);
the color code corresponds to the time in the simulation.
Together, the normalized CR and magnetic pressure encompass 15-20\% of the pressure budget in the postcursor:
$\xi_c$ increases quickly to a value $\gtrsim 0.1$ and remains nearly constant throughout the simulation, whereas the magnetic pressure rises more slowly up to $0.05-0.075$ towards the end of the simulation.
In appendix B of \cite{haggerty+20}, we solve the shock jump conditions between far upstream and the postcursor, including the contributions of CRs and Alfv\'en-like structures in the conservation of mass, momentum, and energy; 
such a solution accounts for the extra-compression observed in the simulation, where $\rt\to\sim 6$, as shown in the bottom left panel of Fig.~\ref{fig:R};
to stress the importance of the postcursor in the shock dynamics, such a panel also includes as a dashed line the prediction with no CR/magnetic drift. 
The CR pressure alone, without the magnetic/drift terms, is not sufficient to account for the strong shock modification that we observe, which demonstrates that the effect is inherently different from the enhanced compression expected in the classical theory of efficient DSA \cite{jones+91,malkov+01}.

\subsection{Steep Spectra}
Drastically different from the classical theory, an enhanced shock compression is not associated to flatter CR spectra, but rather to steeper ones, as shown in the right panels of Fig.~\ref{fig:R}. 
The standard prediction is that the CR momentum spectrum should flatten with time: such expected spectra would be described by Eq.~\ref{eq:DSA} with $r\to \rt$ and are shown with dashed lines in the middle panel.
Instead, the measured postshock CR spectra (solid lines) are systematically steeper than such a prediction.

Since CRs do not feel the change in speed of the thermal plasma, but rather that of their scattering centers, the effective compression ratio that they feel is:
\begin{equation}\label{eq:trt}
    \trt\simeq \frac{u_0}{u_2+\w2}\simeq \frac{\rt}{1 +\alpha}; \quad 
    \alpha\equiv \frac{\w2}{u_2},
\end{equation}
where in the numerator we set $\w0\approx 0$ because upstream infinity fluctuations should be small.
The $\alpha$ parameter quantifies the effect of the postcursor-induced spectral modification; 
since $\alpha>0$ (waves move towards downstream), the compression ratio felt by the CRs is always smaller than the fluid one.
%At the same time, for low-energy CRs that probe the subshock only
%\begin{equation}
%    \trs\simeq \frac{u_1-v_{A,1}}{u_2+v_{A,2}}\simeq \rs\frac{1-\alpha_1}{1 +\alpha}; \quad 
%    \alpha_1\equiv \frac{\w1}{u_1}.
%\end{equation}
Note that, when the magnetic field is compressed at the shock and $B_2\approx\rs B_1$, we have $\alpha= \rs^{3/2}  \alpha_1\lesssim 8\alpha_1$ and the correction due to the postcursor dominates over the one in the precursor.
%This is why we focus on the parameter $\alpha$, which technically might be labelled as $\alpha_2$, rather than on $\alpha_1$.

CR spectra turn out to be even steeper than $p^{-4}$ and match very well the slope calculated using $\trt$, namely
\begin{equation}\label{eq:qtilde}
    \Tilde{q}\equiv \frac{3\trt}{\trt-1}=
    \frac{3\rt}{\rt-1-\alpha}.
\end{equation}
This is plotted as dashed lines in the bottom right panel of Fig.~\ref{fig:R}.
In Bell's approach (Eq.~\ref{eq:bell}), we interpret the steepening as induced by an increase in the escape probability, rather than to a reduction of the acceleration rate (figure 3 in \cite{caprioli+20}).
%We have extensively checked this result against simulation parameters such as number of particles per cell, box transverse size, grid resolution, and time step choice; 
%we rule out that the steep spectra are a numerical artifact to the best of our knowledge.

\section{A Revised Theory of Efficient DSA}\label{sec:revised}
In summary, an interesting non-linear, physics-rich, picture arises. 
Quasi parallel shocks are inherently efficient in injecting thermal ions into the DSA mechanism \cite{caprioli+14a, caprioli+15};
DSA is self-sustaining and the streaming of energetic particles upstream of the shock triggers violent plasma instabilities (especially, the non-resonant one \cite{bell04}), which foster the rapid scattering and energization of CRs.
If no self-regulating effects kicked in, the DSA efficiency would grow uncontrolled, leading to the flatter and flatter spectra envisioned by the standard theory \cite{jones+91, malkov+01}.
Instead, when acceleration efficiency reaches $\sim 10\%$, the associated generation of magnetic field grows and back-reacts on both the shock modification and on the CR spectrum, as discussed above.
The net result of this non-linear chain is that both CR acceleration and $B$ amplification saturate, yielding a DSA efficiency of $\sim 10\%$ and CR spectra mildly steeper than $p^{-4}$.

One important implication is that the spectrum produced by efficient DSA is not universal, but rather depends on the strength of the self-generated  fields.
This adds a novel, crucial, physical constraint when modeling the non-thermal emission from CR sources, as outlined in \S6 of \cite{caprioli+20} and in \cite{diesing+21}.

\subsection{The Case of SN1006}
Particularly interesting is the case of SN1006, which shows a bilateral symmetry defined by the direction of $\vec{B}_0$ \cite{rothenflug+04,bocchino+11}.
X/$\gamma$-ray emission comes from the quasi-parallel (polar caps) regions \cite{gamil+08,giuffrida+22,SN1006HESS}, implying the presence of multi-TeV electrons, while radio emission is more azimuthally symmetric \cite{rothenflug+04}, suggesting the presence of GeV electrons also in oblique regions.
While efficient ion DSA up to multi-TeV energies is consistent with the low polarization and strong synchrotron emission in the polar caps, electron acceleration in quasi-perpendicular regions is likely boot-strapped via SDA, as outlined here, and then can proceed up to GeV energies just because of the interstellar turbulence \cite{caprioli15p,blasi13}.
Whether \emph{multi-TeV} electrons should also be expected in quasi-perpendicular regions is an interesting question that hinges on the longer-term evolution of these systems. 

It is worth noticing that 
\cite{giuffrida+22} used Chandra/XMM X-ray observations to show that the CR-induced shock modification depends on the orientation of the ambient magnetic field, finding convincing evidence that CR acceleration and field amplification are more efficient in quasi-parallel regions and vanish in quasi-perpendicular ones, as originally predicted \cite{caprioli+14a}.
In particular, in quasi-parallel regions the shock compression ratio clearly exceeds $R=4$, by almost a factor of 2;
moreover, the slope of the accelerated particles inferred from synchrotron emission is $q_E\approx 2.3$, steeper than the DSA theory \cite{rothenflug+04}.
Both pieces of evidence, as well as the level of magnetic field amplification inferred from the non-detection of a precursor in the X-rays \cite{morlino+10}, agree perfectly well with the revised DSA theory that hinges on the role of the postcursor in shaping the shock dynamics and the CR spectral slopes.

\section{Beyond Proton Acceleration}\label{sec:heavies}
These notes do not have the presumption to present all the monumental work done by several groups in the past decades to unravel shock acceleration.
The review of the recent progress, which mostly stemmed out from kinetic simulations, is heavily biased towards the theory of proton acceleration, which almost invariably controls the overall shock dynamics.
This does not directly address the acceleration of electrons and of ions heavier than H, which are important to understand the spectrum and the sources of CRs \cite{hoerandel07,dembinski+18,evoli+21,caprioli15,auger17_coll}.

Nevertheless, DSA is general enough that particles with the same rigidity should undergo the same acceleration, so that most of the arguments about spectral slopes and maximum energies should still apply.
The main difference, though, is that particles with different mass/charge ratio may be injected into DSA in a different way, the chief example being the fact that the electron/proton ratio is $\sim 10^{-4}-10^{-3}$ both in Galactic CRs and in SNRs \cite{berezhko+04a,morlino+12, sarbadhicary+17}.

\subsection{Acceleration of Ions with Arbitrary Mass and Charge \label{sec:nuclei}}
Hybrid simulations have been used to study the thermalization, injection, and acceleration of ions with different mass/charge ratios, $A/Z$, in non-relativistic collisionless shocks  \cite{caprioli+17}.

\begin{figure}
\begin{center}
\includegraphics[trim=5px 50px 0px 340px, clip, width=0.8\textwidth]{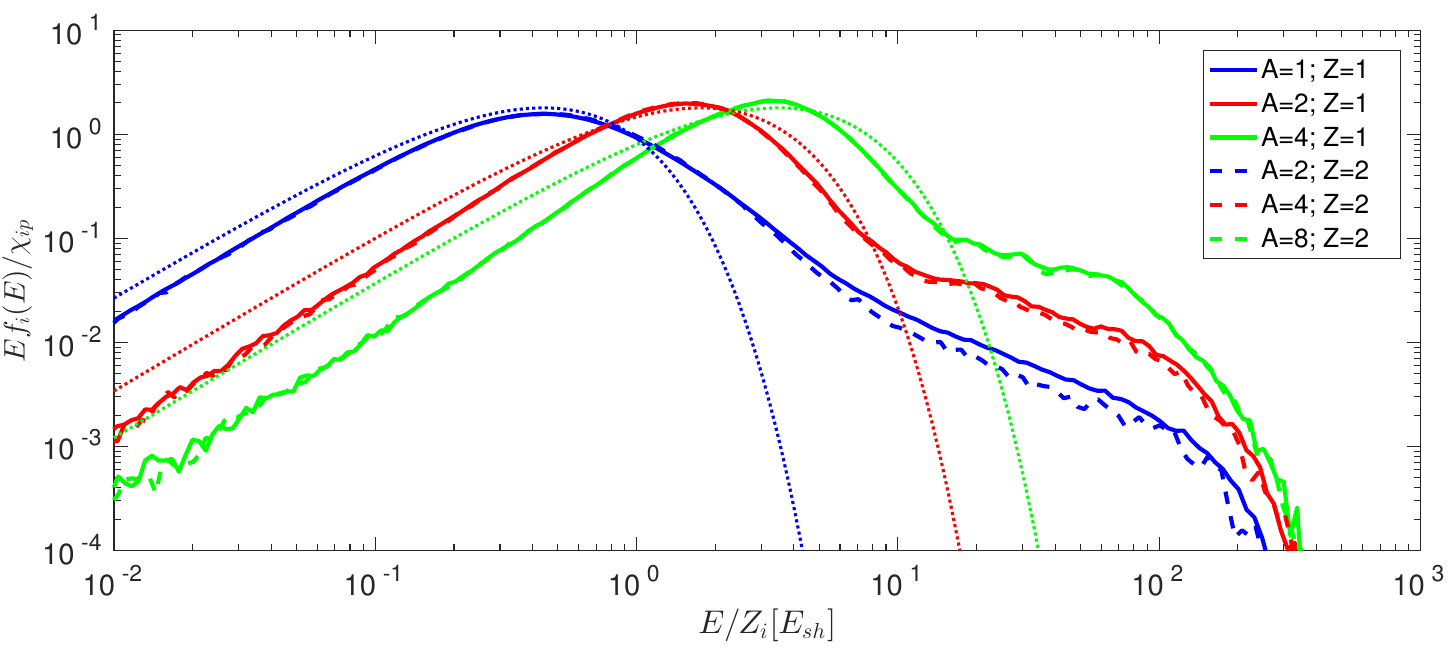}
\caption{Normalized post-shock spectra for ion species with mass $A$ and charge $Z$ as in the legend, for a strong quasi-parallel shock.
Dotted lines (color-matched) correspond to the Maxwellians expected if the temperature scaled with $A$ \cite{caprioli+17}.}
\vspace{-2mm}
\label{fig:spec}
\end{center}
\end{figure}

These results, illustrated in fig. \ref{fig:spec}, can be summarized as follows: 
1) ions thermalize to a post-shock temperature proportional to $A$, which is to be expected since the free kinetic energy available scales with the species' mass;
2) when diffusive shock acceleration is efficient, ions develop a non-thermal tail whose extent scales with $Z$, a manifestation of the fact that DSA is rigidity dependent;
3) the normalization of the power-law tail is enhanced $\propto (A/Z)^2$, so that heavy ions are preferentially accelerated.
This last scaling, never predicted theoretically but just observed in kinetic simulations, provides a quantitative  explanation for the observed chemical composition of Galactic CRs, which are systematically richer in heavier nuclei (see figure 3 in \cite{caprioli+17}). 

The preferential injection and acceleration of heavy nuclei depend on the shock strength: for shocks with lower Mach number the enhancement is less prominent, proportional to $A/Z$.
Moreover, we find that proton and ion injection depends on the shock inclination, being suppressed for magnetized oblique and quasi-perpendicular shocks. 
These two trends hinge on the pivotal role that the self-generated magnetic turbulence has in promoting the injection of heavy nuclei, which need to be heated non-collisionally at the shock crossing (also see \cite{hanusch+19a, hanusch+19b}).

\subsection{Electron Acceleration}
More than a subsection, this part would deserve an entire paper, given all the effort gone into trying to unraveling why electrons are injected into DSA less efficiently than protons. 
Energetic electrons, despite being subdominant in number and energy density with respect to energetic ions, are responsible for most of the non-thermal radiation produced by a shock, emitting from radio to $\gamma$-rays via synchrotron, bremsstrahlung, and inverse-Compton emission.

Hybrid simulations have been pivotal in shaping the current theory of ion DSA and, in general, of CR-modified shocks, but they lack a kinetic treatment for the electrons and cannot address one  outstanding question that arises in modeling multi-wavelength emission from shock-powered systems: \emph{When and how are electrons accelerated}?

Accounting for electron physics naturally requires \emph{full-PIC simulations}, which are dramatically more expensive than hybrid ones:
for the same problem in physical time/lengths, PIC is more expensive by a factor $\propto \mathcal M^{d/2+1}$, where $\mathcal M$ is the proton/electron mass ratio and $d$ is the number of spatial dimensions of the computational box\footnote{This scaling comes from resolving the electron inertial length and plasma frequency rather than the ion inertial length and cyclotron frequency. The problem worsens when also the Debye length must be resolved \cite{shalaby+18}, but this is not usually necessary for shocks \cite{sironi+11}.}. 
A natural choice for keeping simulations manageable is to use an artificial value of $\mathcal M\ll 1836$ and a reduced box dimensionality, but these choices are not harmless because important pieces of the physics may be lost, as it was pointed out by several authors \cite{riquelme+11,xu+20}.

In a nutshell, the processes that promote ion injection (outlined in \S\ref{sec:hybrid}), do not necessarily work for electrons, which have smaller gyroradii and opposite charge;
instead, electrons need to rely on conservation of their magnetic moment \cite{ball+01} and some  pre-acceleration mechanism in order not to be advected downstream and thermalized after crossing the shock. 

From both simulations and observations, we now know that particle injection depends on the shock inclination $\thbn$: 
for $\vartheta\lesssim 45\deg$ (quasi-parallel shocks), ions are spontaneously injected into DSA and electron can be injected thanks to the magnetic turbulence driven by such energetic ions \cite{park+15,crumley+19}; 
however, for  $\vartheta\gtrsim 45\deg$ (oblique shocks), ion injection is suppressed in magnetized shocks \cite{caprioli+15} and electrons have to drive their own waves in order to diffuse back to the shock.
A lot of effort from several groups has gone into unraveling which pre-acceleration process(es) may eventually lead to electron injection into DSA for different simulations parameters;
a non-comprehensive list limited to the most promising ones includes: whistler waves, oblique firehose modes, shock surfing acceleration, shock drift acceleration, intermediate-scale instability, electron-cyclotron drift instability, and magnetic reconnection in the shock foot
\cite{lembege02a,krasnoselskikh95a,lembege+04,amano+07,amano+09a, amano+09b,amano+10,guo+14a,guo+14b,savoini+13,matsumoto+15,matsukiyo+06, bohdan+19a,bohdan+19b,bohdan+20a,bohdan+20b,marcowith+16,shalaby+22,fiuza+20,wang+22,morris+22,morris+23, shimada+00,ha+21,ha+22,ha+23,kang+19,kim+21}.
While these papers demonstrate that modeling electron DSA from first principles is possible, a comprehensive theory of electron injection and acceleration is still missing (see, e.g., the very recent review \cite{amano+22}).

\section{PIC Simulations of Non-relativistic Shocks\label{sec:pioneer}}
\begin{figure}
\centering
\vspace{-5mm}
\includegraphics[trim=0px 0px 0px 0px, clip=true, width=0.44\textwidth]{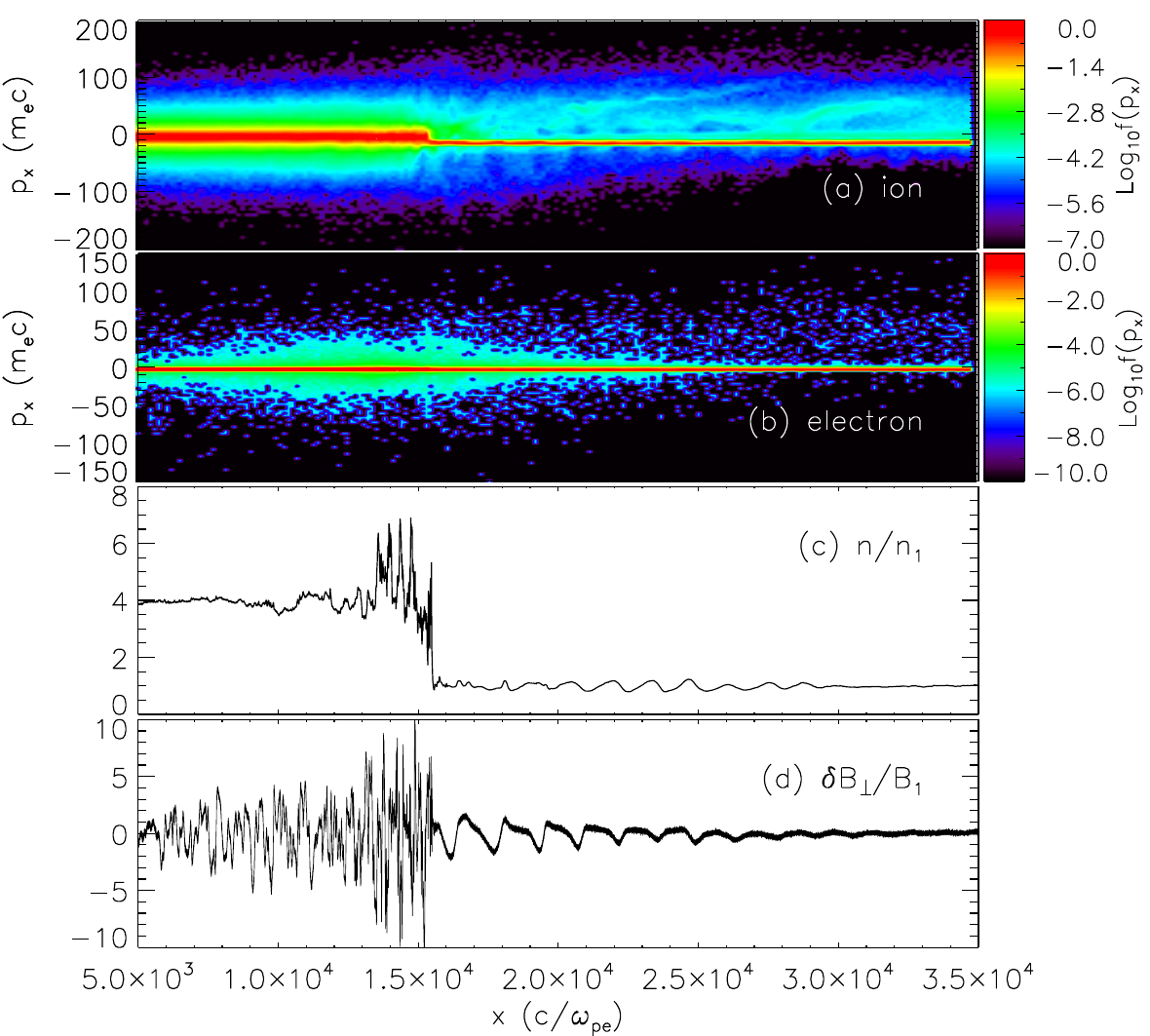}
\includegraphics[trim=0px 0px 20px 0px, clip=true,width=0.54\textwidth]{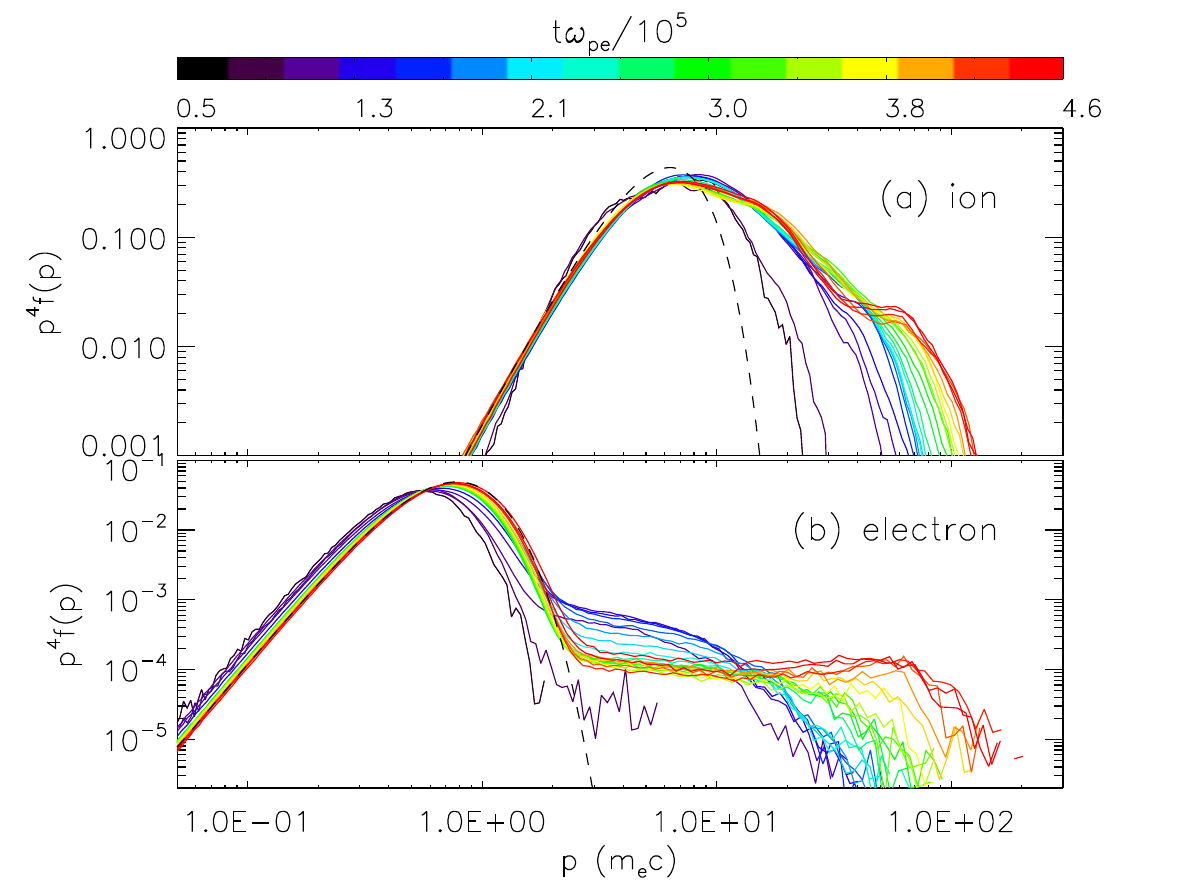}
\caption{\label{fig:ele} 
\emph{Left Panel:} Phase space distributions of ions (a) and electrons (b), density (c), and transverse component of the B-field (d) for a PIC simulation of a non-relativistic quasi-parallel shock.
Energetic particles diffuse ahead of the shock, amplifying the upstream magnetic field.
\emph{Right Panel:} Evolution of the post-shock spectra for (a) ions and (b) electrons; note that both species develop a power-law tail \cite{park+15}.
}
\end{figure}

As mentioned above, the shock inclination controls ion injection into DSA; it also plays a fundamental role in electron injection, so we will distinguish between quasi-parallel and oblique shocks. 
Moreover, for every shock velocity there exists a critical angle $\thbn^*$ above which the shock becomes \emph{superluminal}, i.e., even particles moving along a magnetic field line at the speed of light cannot overrun the shock because the projection of their velocity along the shock normal is invariably smaller than the shock speed \cite{sironi+09}.
Such a condition can be expressed as $\cos{\thbn^*}=\vsh/c$, so the problem of having particles escaping upstream (necessary to undergo DSA) is more serious at trans-relativistic and relativistic shocks.

One may expect that the direction of the magnetic field matters less and less if the shock magnetization becomes smaller and smaller, i.e., if the shock Alfv\'enic Mach number is sufficiently large. 
Qualitatively, one can also think that if the plasma $\beta\equiv P_{\rm th}/P_{\rm B}\approx M_A^2/M_s^2$ (i.e., the ratio of the upstream thermal to magnetic pressure, $M_s$ being the sonic Mach number) is large enough, particles are less magnetized and more susceptible to be injected into DSA. 
Therefore, we expect that both ion and electron acceleration should depend on four main parameters: the shock inclination $\thbn$, the shock speed $\vsh/c$, and the Alfv\'enic and sonic Mach numbers, $M_A$ and $M_s$ (or equivalently $\beta$).

On top of these parameters set by the shock environment, there are additional parameters that need be adjusted to keep the simulation numerically acceptable. 
Such parameters are the reduced proton/electron mass ratio, $\mathcal{M}\lesssim 1836$, and the dimensionality of the simulation box. 
Time/space resolution and number of particles per cell are further model parameters that require a convergence test \cite{shalaby+17}.

\subsection{Quasi-parallel shocks}
\begin{figure}[t]
\centering
\includegraphics[trim=30px 10px 0px 10px, clip=true, width=0.5\textwidth]{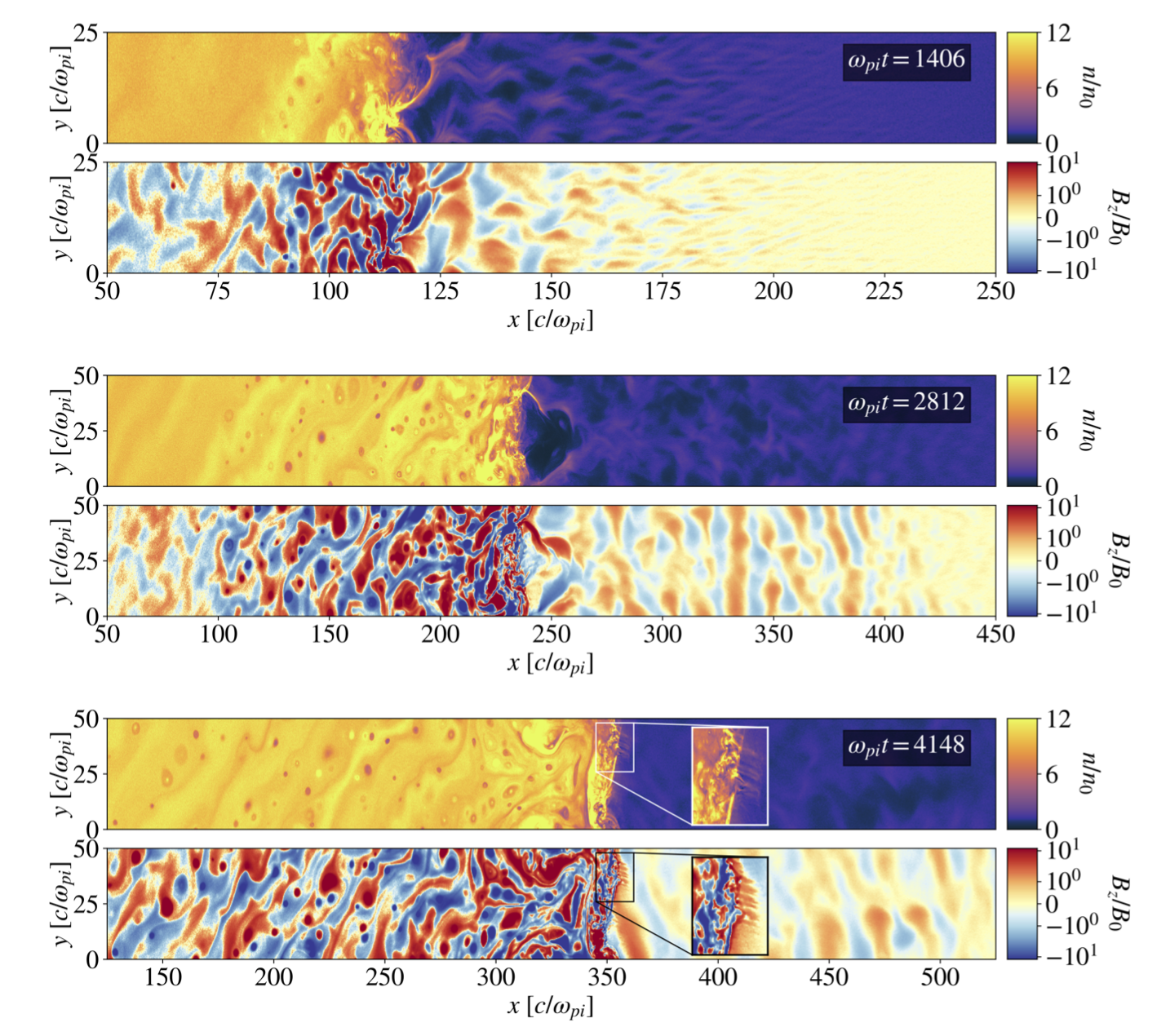}
\includegraphics[trim=5px 3px 10px 7px, clip=true, width=0.45\textwidth]{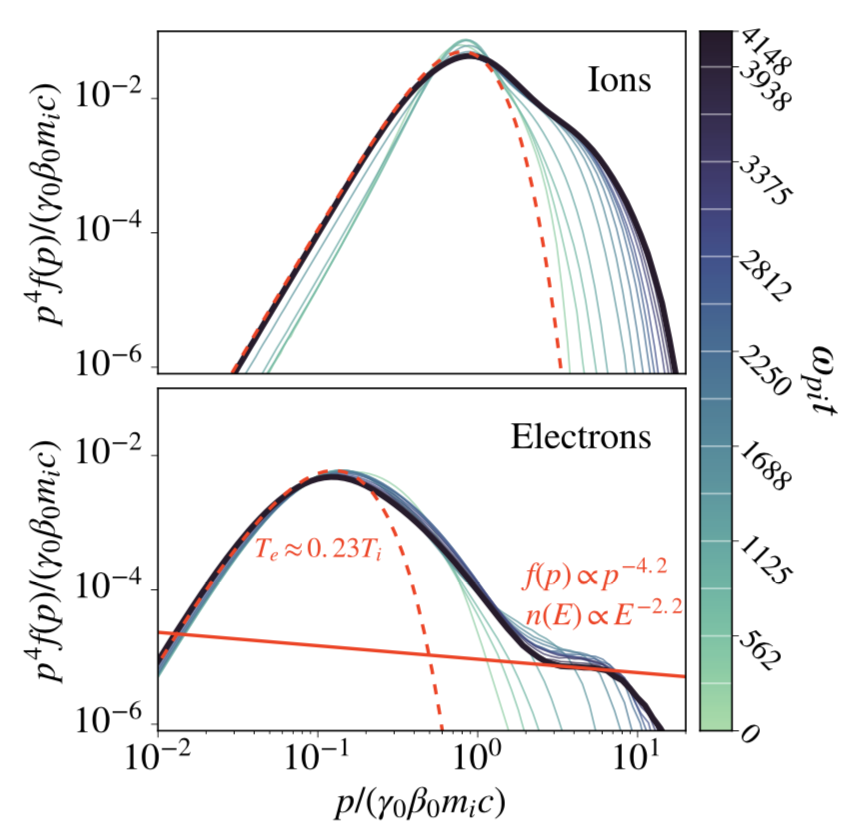}
 \caption{\label{fig:PIC}
2D PIC simulation of a trans-relativistic quasi-parallel shock.
 \emph{Left panels}: Time evolution of plasma density $n$ and magnetic field $B$ (top to bottom); note the turbulence that is entirely generated by accelerated particles.
\emph{Right panel}: Time evolution of particle non-thermal tails $\propto E^{-2.2}$. From \cite{crumley+19}.}
\vspace{-15pt}
\end{figure}

In \cite{park+15} the first PIC simulation that showed the simultaneous development of DSA tails in both electron and ion distributions at a non-relativistic shock was put forward (Fig.~\ref{fig:ele}).
In order to capture the diffusion lengths of accelerated ions, and the instability that they produce, a very long box (about 10 million cells, a few times $10^4d_i$, where $d_i$ is the ion skin depth) is needed, which makes it challenging to go beyond 1D. 
The faster the shock, the closer all the velocity scales are, which results in a lower computational burden.
This allowed us to successfully run trans-relativistic shocks with $\vsh=0.3-0.8c$ \cite{crumley+19} even in 2D, confirming the simultaneous acceleration of electrons and ions (Fig.~\ref{fig:PIC}).
The overall picture that arises from these simulations is that at quasi-parallel shocks both species are spontaneously injected into DSA and accelerated to larger and larger energies.
The process is \emph{self-similar}: the maximum ion energy increases with time because particles scatter on top of the magnetic perturbations that they generate upstream.

\subsection{Oblique shocks}
\begin{figure}[t]
\centering
\vspace{-15pt}
\includegraphics[trim=0px 0px 0px 0px, clip=true, width=\textwidth]{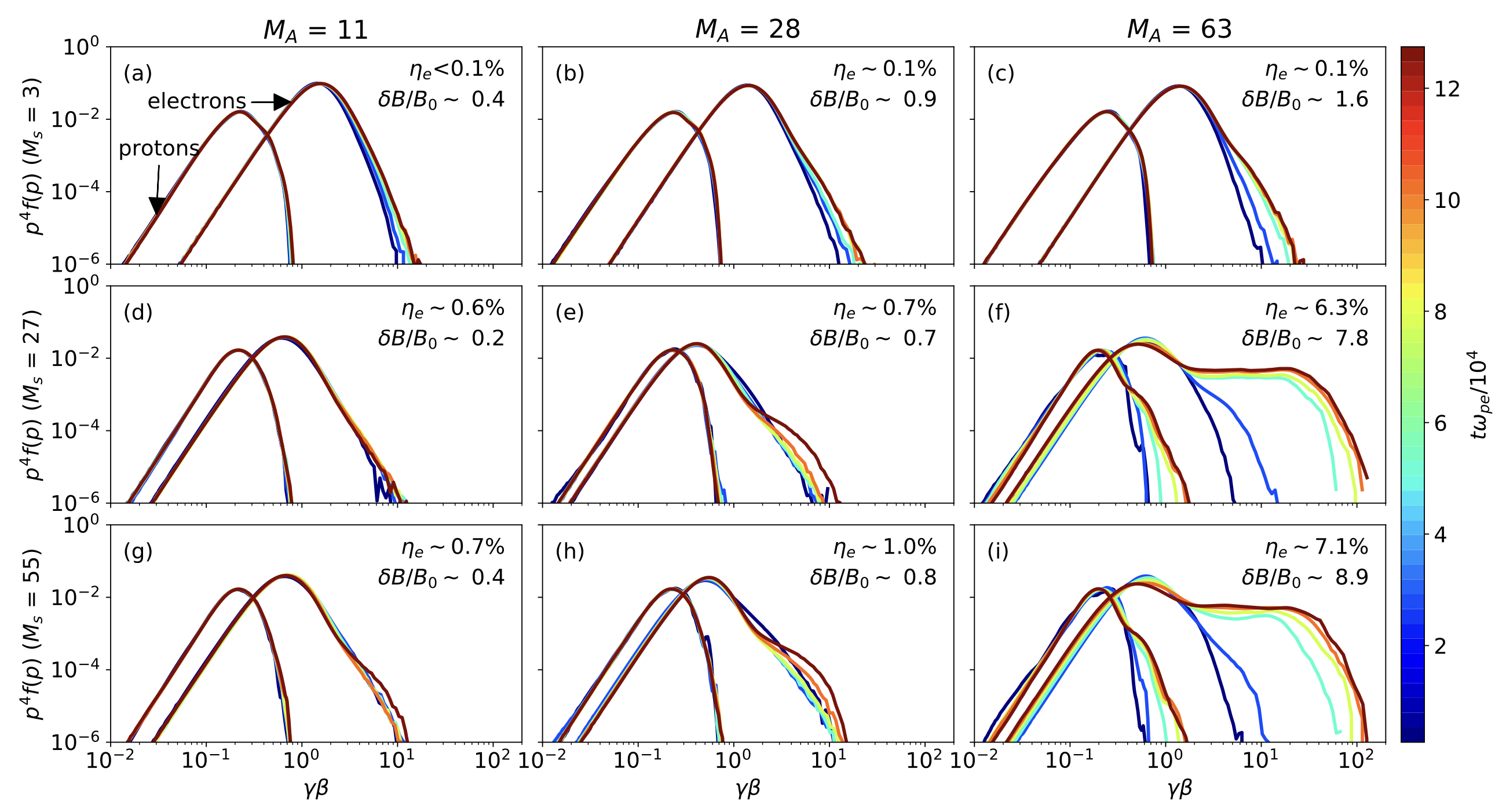}
\vspace{-3mm}
 \caption{\label{fig:oblique1D}
Time-dependent spectra obtained in 1D PIC simulations of oblique shocks ($\thbn=63\deg$) for different sonic and Alfv\'enic Mach numbers as indicated. The electron acceleration efficiency and the amplification of the upstream magnetic field are reported in the legends.
 Note the trend of more efficient electron acceleration with increasing Mach numbers. From \cite{xu+20}.}
\end{figure}

In \cite{xu+20} a 1D survey of electron and ion acceleration at oblique shocks with $\thbn=63\deg$ has been performed.
By exploring the parameter space of different sonic and Alfv\'enic Mach numbers, they found that high Mach number quasi-perpendicular shocks can efficiently accelerate electrons to power-law spectra (Fig.~\ref{fig:oblique1D}). 
Electrons are reflected by magnetic mirroring at the shock and drive non-resonant waves in the upstream. 
Reflected electrons are trapped between the shock front and upstream waves, and undergo multiple cycles of shock drift acceleration before being injected into DSA. 
Strong current-driven waves also temporarily change the shock obliquity and cause mild proton pre-acceleration even in quasi-perpendicular shocks, which usually have hard time accelerating protons. 
The injection of electrons in high-$\beta$, weak ($M_s\lesssim 5$), oblique shocks has also been investigated in 2D \cite{guo+14a,guo+14b,ha+21,kang+19,ha+22,ha+23}; 
such an environment seems to favor electron injection into DSA, even if conclusive evidence of DSA has not been reported, yet. 

\subsection{Relativistic Shocks}
Here we refer to the pioneering PIC simulations performed by A. Spitkovsky and L. Sironi \cite{spitkovsky08,sironi+09, sironi+11,sironi+13}, which cover the parameter spaces that lead to DSA of electron, positrons, and ions. 
In a nutshell, when all the species' velocities become very close to $c$, there is little difference between electrons and protons, in the sense that they are injected into DSA in a similar way.

Spectra tend to be slightly steeper than the test-particle prediction because particles are advected away downstream more effectively than in non-relativistic shocks (i.e., particles are not isotropic in the downstream \cite{blasi-vietri05, morlino+07a, morlino+07b}), as discussed in the thorough review by \cite{jones+91}.
Note that PIC simulations have not been pushed long enough to see non-linear DSA in action, so this may need to be revised, too.

Another main difference with respect to non-relativistic shocks is that the turbulence driven by accelerating particles seems to be too small-wavelength to efficiently scatter particles.
At non-relativistic shocks $E_{\rm max}\propto t$ \cite{park+15}, while at relativistic shocks generation of turbulence is hindered by the shorter advection timescales and acceleration is slower, $E_{\rm max}\propto \sqrt{t}$ \cite{sironi+13}.

\section{Conclusions}
In these notes we introduced the fundamentals of shock acceleration (Fermi mechanisms, shock hydrodynamics, DSA) and the general expectations that stem from the linear theory (\S\ref{sec:intro}).
We proceed by discussing the limitations of such a theory and the pieces of observational evidence that complement and challenge it (\S\ref{sec:challenges}).

Then, in \S\ref{sec:sims} we presented the modern kinetic hybrid (kinetic ions---fluid electrons) plasma simulations that validate and complete the predictions for the spectral slope of the accelerated particles and quantify the most elusive ingredients that a linear theory cannot predict (injection efficiency, maximum achievable energy).
Unprecedentedly-long hybrid simulations are discussed in \S\ref{sec:modified}, where the main deviations from linear theory are introduced, most notably the formation of a post-cursor, which modified both the shock compression ratio and the spectrum of the accelerated particles.

A revised theory of DSA arises (\S\ref{sec:revised}), which agrees well with observations of many shock-powered astrophysical objects, in particular SNRs.

Finally, we outline the progress in unraveling the injection and acceleration of heavy ions in \S\ref{sec:heavies} and electrons (\S\ref{sec:pioneer}), for which PIC simulations are crucial players.

We conclude by pointing out what the author believes are the most important missing pieces of a complete theory of shock acceleration:
\begin{itemize}
    \item understanding electron injection as a function of the shock parameters; this is a key ingredient (and currently essentially a free parameter) in modeling the multi-wavelength emission from shock-powered astrophysical systems \cite{corso+23};
    \item quantifying the maximum energy achievable as a result of the CR-driven instabilities; this encompasses characterizing the saturation of the Bell instability \cite{bell04,zacharegkas+22} in the context of Galactic accelerators \cite{bell+13,blasi+15,cristofari+21,cristofari+22};
    \item understanding the long-term evolution of oblique and quasi-perpendicular shocks \cite{orusa+23}; this requires full-3D simulations, but has the potential to address the detailed phenomenology of many space/astro shocks.
\end{itemize}

\acknowledgments
These notes cover the introduction to shock acceleration presented at the International School of Physics ``Enrico Fermi" on \emph{Foundations of Cosmic Ray Astrophysics}, held in Varenna (Italy), in June 2022.
I hope this may be useful to the next generation of cosmic rays astrophysicists!

I warmly thank my mentors Pasquale Blasi, Mario Vietri, and Anatoly Spitkovsky for their guidance, competence and passion, and for having introduced me to cosmic ray and astroplasma physics. 
I also want to acknowledge how much I have learned standing on the shoulders of the founders of the physics in these notes (in alphabetical order): A. Bell, R. Blandford, L. O'C. Drury,  D. Ellison, D. Eichler, T. Gaisser, T. Jones, M. Malkov, and W. Matthaeus. 
Last, but not least, all of my collaborators in the papers mentioned here, especially E. Amato, S. Gupta, C. Haggerty,  B. Schroer, L. Sironi, L. Wilson III, and my students R. Diesing,  R. Mbarek, L. Orusa, E. Simon, and G. Zacharegkas.

Simulations were performed on computational resources provided by the University of Chicago Research Computing Center.
D.C. was partially supported by NASA through grants 80NSSC20K1273 and 80NSSC18K1218 and NSF through grants AST-1909778, PHY-2010240, and AST-2009326.

\bibliography{Total.bib}
\bibliographystyle{varenna}

%\begin{thebibliography}{0}
%\bibitem{ref:apo} \BY{Einstein A. \atque Fermi E.}
%  \IN{Phys. Rev. A}{13}{1999}{12};
%  \SAME{69}{999}{1666}.
%\end{thebibliography}

\end{document}